\documentclass[preprint,showpacs,preprintnumbers,amsmath,amssymb,superscriptaddress,prb]{revtex4}  
\usepackage{amsbsy}
\usepackage{color}
\usepackage{graphicx}
\usepackage[hyperindex,breaklinks,linktocpage,colorlinks]{hyperref}   

\def\bsym#1{\boldsymbol{#1}}
\def\d#1{{\rm d}#1}
\def\eps{\varepsilon}
\def\eV{{\rm eV}}
\def\gdir#1{\langle #1 \rangle}
\def\GPa{{\rm GPa}}
\def\k{\bsym{k}}

\def\m{{\rm m}}

\def\rr{\bsym{r}}
\def\refapx#1{Appendix~\ref{#1}}
\def\refeq#1{(\ref{#1})}
\def\reffig#1{Fig.~\ref{#1}}
\def\reftab#1{Table~\ref{#1}}
\def\sym{{\rm sym\ }}
\def\um{{\rm \mu m}}

\begin{document}

\title{Defect-induced incompatibility of elastic strains: dislocations within the Landau theory of
  martensitic phase transformations}

\author{R. Gr\"oger}
\email{groger@lanl.gov}

\author{T. Lookman}
\author{A. Saxena}
\affiliation{Theoretical Division and Center for Nonlinear Studies, Los Alamos National Laboratory, 
  Los Alamos, NM 87545, USA}

\begin{abstract}
  In dislocation-free martensites the components of the elastic strain tensor are constrained by the
  Saint-Venant compatibility condition which guarantees continuity of the body during external
  loading.  However, in dislocated materials the plastic part of the distortion tensor introduces a
  displacement mismatch that is removed by elastic relaxation. The elastic strains are then no
  longer compatible in the sense of the Saint-Venant law and the ensuing incompatibility tensor is
  shown to be proportional to the gradients of the Nye dislocation density tensor.  We demonstrate
  that the presence of this incompatibility gives rise to an additional long-range contribution in
  the inhomogeneous part of the Landau energy functional and to the corresponding stress
  fields. Competition amongst the local and long-range interactions results in frustration in the
  evolving order parameter (elastic) texture.  We show how the Peach-Koehler forces and stress
  fields for any distribution of dislocations in arbitrarily anisotropic media can be calculated and
  employed in a Fokker-Planck dynamics for the dislocation density. This approach represents a
  self-consistent scheme that yields the evolutions of both the order parameter field and the
  continuous dislocation density. We illustrate our method by studying the effects of dislocations
  on microstructure, particularly twinned domain walls, in an Fe-Pd alloy undergoing a martensitic
  transformation.
\end{abstract}

\date{\today}

\pacs{81.30.Kf, 63.70.+h, 61.72.Lk, 05.70.Fh, 05.10.Gg}

\keywords{martensitic transformation; Landau-Ginzburg; dislocations; plasticity; Saint-Venant law;
  incompatibility}

\maketitle


\section{Introduction}

A mesoscopic description (nano to micrometer) of physical processes in solids, where atomic length
scales merge with those of the continuum, represents a crucial and perhaps most challenging aspect
of understanding material behavior. This arises, for example, during displacive (martensitic) phase
transformations where the distortions associated with the strains in unit cells and intra-unit cell
displacements (or shuffles) propagate over larger distances so that competing long-range effects
lead to the formation of inhomogeneities such as interfaces, spatially correlated domains and
complex microstructure.  It is the least understood regime compared to the atomic and continuum
scales because the simplifications and advantages of theory in handling small/large length scales
and fast/slow time scales no longer apply.  The predictions of atomistic models become typically
invalid at length scales larger than about a nanometer, whereas the lack of the detailed description
in continuum theories makes them inappropriate for studies of physical processes occurring at length
scales below a millimeter.  

Our focus will be on materials undergoing structural phase transformations that represent an
important and broad class of advanced materials frequently utilized in state-of-the-art applications
such as surgical tools, artificial muscle fibers, aerospace and robotic applications, and novel
microelectronic devices.  Examples include shape memory alloys (Ni-Ti, Fe-Pd, Au-Cd, Cu-Al-Ni),
nuclear materials (Pu, U-Nb), ferroelectrics with spontaneous polarization (BiFeO$_3$,
LuFe$_2$O$_4$, LiCu$_2$O$_2$), strain-induced ferroelectric perovskites and high-k insulators
(BaTiO$_3$, SrTiO$_3$, PbTiO$_3$, LaAlO$_3$), magnetic shape memory alloys (Fe-Pd, Ni$_2$MnGa), or
even materials with ferrotoroidic ordering (LiCoPO$_4$, Co$_3$B$_7$O$_{13}$Br(I)).  Related
materials with additional degrees of freedom include the recently discovered single phase
multiferroics displaying magnetoelectricity (BiMnO$_3$, TbMnO$_3$, HoMnO$_3$) where magnetization
and polarization are coupled to the lattice and where noncollinear structures arise. The stable
crystal structure in martensites at high temperatures, often of cubic symmetry, is identified as
\emph{austenite}. Near the transformation temperature $T_c$, the coordinated motion of all atoms
spontaneously breaks the symmetry of the high-temperature phase and several variants of the
low-temperature phase, the \emph{martensite}, ensue.  This phase is a combination of all the
individual martensite variants, the fraction of which in the microstructure depends on many factors
such as the ambient temperature, rate of cooling, or applied stress. The typical symmetries of the
martensite include tetragonal (Fe-Pd, In-Tl, Ni-Al(Mn), Fe-Ni-C, BaTiO$_3$, Ni$_2$MnGa),
orthorhombic (Au-Cd, U$_6$Nb, Ti-Ta(Pd), Cu-Ni-Al(Ti), U), monoclinic (Ni-Ti, Ni-Ti-Al(Cu,Fe,Mn,Pd),
Cu-Al-Zn(Be), Zr$_2$CuCo, Pu), or trigonal/rhombohedral (Au-Cd, Ni-Ti, Ti-Ni-Al(Fe)). As recognized
early by Landau \cite{landau:36, landau:37a, landau:37b}, an important condition for displacive
phase transformations to occur is that the symmetry group of the martensite is a subgroup of the
symmetry group of the austenite.  The ratio between the number of the symmetry operations of the
austenite and the martensite then determines the number of different martensite variants possible in
the microstructure below $T_c$. For example, for a cubic to monoclinic phase transformation the
initial cubic symmetry is described by 48 symmetry elements whereas the final monoclinic symmetry by
only 4.  Below $T_c$, the high-temperature cubic symmetry is thus spontaneously broken into 12
variants of the martensite.  Because all martensite variants have the same energy, the final
microstructure below $T_c$ is typically composed of regions accommodating these individual variants.
When different martensite variants are brought together to form an interface, it leads to a
strain-matched or strain-free twin boundary or domain wall, if properly oriented.  Otherwise, there
exist transition zones at the domain boundaries that contribute extra compressional and shear
energies, and it may thus be favorable for the structure to form twin boundary dislocations to lower
the free energy. Similarly, a habit plane between the parent austenite and product martensite
variants is an invariant strain plane if properly matched and oriented so that there is a twinned
microstructure in the martensite with rapidly decaying strain fields in the austenite. Upon cooling
from the austenitic phase, the individual martensite variants may form a precursor or tweed-like
microstructure just above the transition temperature which then transforms into a fully twinned
pattern below the transformation temperature. These morphologies have been well characterized
experimentally, for example, in the L$1_0$ $\gamma_1$-phase of the intermetallic Fe-Pd
\citep{gushchin:84, greenberg:03, xu:04}, Ni-Ti \citep{sehitoglu:03, daly:07}, Au-Ti
\citep{inamura:07}, U-Nb \cite{field:01}, and Cu-Zn-Al \cite{gall:98}.

Energy minimizing principles are now widely employed to study equilibrium microstructure and
evolution of martensitic phase transformations. The approach pioneered by Barsch and Krumhansl
\cite{barsch:84} utilizes a nonlinear free energy together with strain inhomogeneity in terms of
appropriate order parameter strains that drive the transformation from austenite to martensite. The
one-dimensional interface solutions were shown to be soliton-like and part of the motivation was to
demonstrate how martensite formation can be described within a continuum framework without the need
for invoking dislocations. These ideas were subsequently extended to two dimensions, however, the
solutions for the martensitic structure were always in terms of displacement fields \cite{jacobs:95}
rather than strains. The effects of compositional fluctuations and a description of the precursor,
the tweed microstructure, in Fe-Pd was considered in a Monte Carlo study of \citet{kartha:95} in
which the free energy was written in terms of strains that contained elastic signatures associated
with compositional effects.  The strains were written as gradients of the displacement field and the
Saint-Venant compatibility constraint served as the integrability condition for strain fields.
Incorporating this constraint leads to long-range interactions in the order parameter fields and the
minimum of the free energy is obtained for a twinned microstructure typical of the martensite phase.
The multiscale consequences of the strain-only model were investigated by \citet{shenoy:99} and
the nature of the repulsive potential associated with the microstructure, strain order parameter
dynamics and extensions to two-dimensional ferroelastic transformations are reviewed by
\citet{lookman:03} Applications to phase transformations in shape memory polycrystals and dynamic
strain loading in martensites undergoing a cubic to tetragonal transformation were further studied by
\citet{ahluwalia:06}  The validity of the Saint-Venant compatibility constraint guarantees that the
strain field can be obtained from a known displacement field by taking gradients. On the other hand,
if one knows the strain field, the displacement field is determined up to a rigid body motion by
integration. These statements are only true when the material does not contain any topological
defects. For example, if dislocations are present the displacement field becomes multivalued and its
gradient, i.e.  the strain field, is not defined. Hence, previous studies using a strain-only
description are valid for defect-free media only.

Mesoscopic studies of the collective phenomena associated with defects have focused primarily on the
mechanisms involved in dislocation pattern formation. Groma et al. \cite{groma:97, groma:99} and
Bak\'o et al.  \cite{bako:99a} formulated a statistical model for the evolution of the dislocation
density in isotropic bodies, where the stress field associated with each dislocation is given
analytically \cite{hirth:82, hull:01}. For the case of single slip in two dimensions, the
dislocation density evolves according to a Fokker-Planck equation in which the Peach-Koehler force
\cite{peach:50} on each dislocation is determined from the known stress field. A different
framework, based on statistical studies of dislocation patterning developed by El-Azab
\cite{el-azab:00,el-azab:00a}, makes closer connection with the Kr\"oner's \cite{kroner:58}
continuum theory of dislocations.  It accounts not only for long-range interactions between
dislocations but also treats each discrete slip system separately. By calculating pair correlations
between dislocations in this statistical model, Zaiser et al.\cite{zaiser:01} demonstrated that
dislocation systems exhibit a patterning instability and this leads to the formation of dislocation
walls perpendicular to the glide plane. These statistical ensembles of dislocations have been shown
to exhibit intrinsic spatio-temporal fluctuations with scale-invariant characteristics, long-range
correlations and emergence of strain bursts (for a recent review, see Zaiser \cite{zaiser:06}). In
addition, continuum theories of dislocations and self-stresses as developed by \citet{kroner:58},
\citet{kosevich:64} and others have been applied to studies of dislocation patterning. Among the
most prominent are contributions of \citet{kratochvil:03} and \citet{sedlacek:07} where the
evolution of the dislocation density is formulated in both the Eulerian and the Lagrangian frames. A
phase field formulation of the dislocation patterning in isotropic media was developed by
\citet{rickman:97}, where the dislocation density tensor is obtained by minimizing the free
energy. However, this model does not include coupling between the dislocation density and the
underlying crystal structure and, therefore, it does not exhibit any structural phase transition.
Most recently, \citet{roy:05} implemented the Kr\"oner's theory to study dislocation patterning
using the Finite Element Method (FEM). A similar approach was adopted by \citet{limkumnerd:06} to
formulate a mesoscopic Landau theory in which the free energy is written in terms of the plastic
distortion tensor. The plastic distortion field that minimizes the free energy is then used to
calculate the dislocation density which plays the role of the order parameter. The added advantage
of this formulation is that the same model applies to dislocation motion by pure glide and by a
combination of glide and climb and, therefore, it allows for studies of dislocation patterning in
isotropic materials at both low and high temperatures. However, the stress associated with
individual dislocations is still calculated using the isotropic elasticity \cite{hirth:82, hull:01}.

Our objective in this paper is to incorporate dislocations into the Landau theory to study
martensitic phase transformations in materials containing defects.  Unlike the previous studies
cited above, we consider an anisotropic medium that is described by the elastic constants
corresponding to the high-temperature cubic phase. Utilizing Kr\"oner's \cite{kroner:58} continuum
theory of dislocations, we show that the presence of dislocations induces \emph{incompatibility}
between the elastic components of the strain tensor field, and this is connected to the gradients of
the dislocation density.  The presence of dislocations is responsible for a nonlocal coupling of the
incompatibility field with the order parameter and, as a consequence, the evolving martensitic
texture is affected by the finite density of dislocations.  Minimizing the free energy subject to
the incompatibility constraint for a given distribution of dislocations generates a stress field
that corresponds exactly to this distribution of defects. By inserting a single edge dislocation
into an otherwise ideal crystal we show that the order parameter field that minimizes the free
energy subject to this incompatibility constraint yields the correct long-range stress field around
this dislocation. The fact that the stress field in a generally anisotropic material with arbitrary
distribution of dislocations can be calculated by merely minimizing the free energy means that we
can easily calculate the Peach-Koehler forces that act on these individual dislocations.  These
forces are then used in the Fokker-Planck equation for an evolution of the dislocation density.  The
procedure outlined above represents a self-consistent scheme that is solved recursively. In the
first step, the order parameter field is calculated by minimizing the free energy subject to the
incompatibility constraint that is obtained from the given dislocation density. For the known order
parameter field, the Peach-Koehler forces on individual dislocations are calculated and utilized in
the evolution equation to update the dislocation density and thus also the corresponding
incompatibility field.

The mesoscopic nature of our approach makes this work different from the more nanoscale-based
models, in particular, the phase field microelasticity theory of \citet{khachaturyan:83} and
\citet{wang:01}.  In their work the dislocation loops are viewed as coherent platelet inclusions
that expand, interact with other loops in the same and other slip systems and annihilate in response
to their internal long-range strain fields and externally applied load. The dislocation content of
each slip system is described by an integer-valued density function $\eta$ that specifies the number
of dislocations with prescribed direction of the Burgers vector. The total energy is written in
terms of stress-free strain due to individual dislocation loops and the evolution of the order
parameter field $\eta$ is studied using a Langevin dynamics based on the time-dependent
Ginzburg-Landau equation. This model has been successfully utilized in studies of heterogeneous
nucleation of martensite in the parent austenitic martix (for an excellent review, see
\citet{malygin:01}). The martensitic embryo was shown to grow inside the dislocation loop that
expands in response to the external load, giving rise to the so-called stress-induced martensite. In
contrast, within the Barsch-Krumhansl formulation of martensite \cite{barsch:84}, which is utilized
in this paper, twins nucleate spontaneously by lowering the temperature and without the need of
an externally applied load provided the system contains a certain minimum degree of strain
inhomogeneities induced by thermal fluctuations of the lattice. In the present work, this
is no longer needed because the heterogeneous nucleation of the martensite takes place
readily on preexisting dislocations.

The plan of our paper is as follows. In Section \ref{sec_Kroner} we review the continuum theory of
dislocations and show how we include the dislocation density as the source of the
incompatibility. In Section \ref{sec_freeE} we eliminate this constraint in favor of long-range
interactions in the dislocation density and its nonlocal coupling with the elastic strains.  In
Sections \ref{sec_relax_e2} and \ref{sec_relax_alpha} we show how the total free energy is minimized
using relaxational dynamics and how the dislocation density is evolved using the Fokker-Planck
equation, respectively. In Section \ref{sec_simul} we utilize the theory developed in this paper to
study martensitic phase transformations and dislocation patterning in single crystals of
Fe-30at.\%Pd alloys. In the first case study we consider that the dislocation density is fixed and
thus the free energy is minimized solely by the order parameter field. We demonstrate that the
increase of the dislocation density induces long-range internal strains in the material, and these
give rise to stress-induced martensite even above the temperature $T_c$ for a defect-free
material. Below $T_c$ the morphology of the martensite changes from well-defined $\gdir{110}$ twins
at low dislocation density to a twin-free order parameter field at high dislocation density. In the
second case we consider ideally mobile crystal dislocations. We show that below $T_c$ the order
parameter field evolves into a series of twins corresponding to the two variants of the
martensite. The equilibrium dislocation density is characterized by regions of high dislocation
density in which the dislocations are arranged into walls at twin boundaries, separated by regions
of low dislocation density. We show that the neighboring dislocation walls are formed by opposite
Burgers vectors perpendicular to the twin boundaries. These correlated domains have been observed
not only in molecular dynamics and Monte Carlo simulations\cite{gulluoglu:89, groma:99}, but also in
experiments on Ag\citep{worzala:67}, Ni-Ti\cite{sehitoglu:03} and Fe-Pd\cite{halley:02}.


\section{Continuum theory of dislocations}
\label{sec_Kroner}

\begin{figure}[!htb]
  \centering
  \includegraphics[scale=1.5]{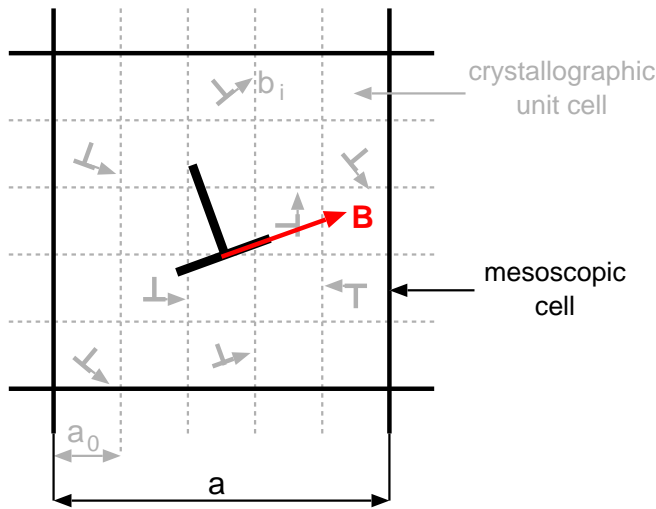} \\
  \caption{Each mesoscopic cell (solid square) is comprised of a finite number of crystallographic
    unit cells (dotted gray squares). The net Burgers vector $\bsym{B}$ is a vector sum of the
    Burgers vectors of individual crystallographic dislocations (gray arrows).}
  \label{fig_mesocell}
\end{figure}

In order to demonstrate how the presence of defects breaks the single-valuedness of the displacement
field, we will first perform the following thought experiment as proposed originally by
\citet{kroner:58,kroner:81}. Consider a macroscopic single crystal whose structure belongs to a
well-defined space group and its unit cell is defined by the lattice parameters $\bsym{a_0} =
\left\{ a_0, b_0, c_0 \right\}$ and angles $\bsym{\alpha_0} = \left\{ \alpha_0, \beta_0, \gamma_0
\right\}$. Only those microscopic details associated with the crystal structure that manifest
themselves at the mesoscopic level will be taken into account. This leads us to discretize the
medium into a finite number of \emph{mesoscopic} cells with lattice parameters $\bsym{a} = \left\{
  a, b, c \right\}$ and angles $\bsym{\alpha} = \left\{ \alpha, \beta, \gamma \right\}$, where each
such cell includes a finite number of the crystallographic unit cells. The dislocation content of
each mesoscopic cell can thus be characterized by the so-called net Burgers vector $\bsym{B}$ that
is defined as a vector sum of the Burgers vectors $\bsym{b}_i$ of the crystal dislocations in all
embedded crystallographic unit cells. This is shown schematically in \reffig{fig_mesocell}. Each of
these ``net dislocations''\footnote{The term ``net dislocation'' is used here in a loose sense and
  should not be thought of as a crystal dislocation. The reason is that the net Burgers vector is a
  vector sum of the Burgers vectors of individual crystallographic dislocations and, therefore, it
  can assume in general any orientation and magnitude.}, characterized by its Burgers vector
$\bsym{B}$, causes a certain plastic distortion $\bsym{\beta^p}$ that is proportional to the
magnitude of $\bsym{B}$.  We require that the continuity of the body be maintained for any
distribution of the net Burgers vectors, i.e.  for any corresponding plastic distortion
$\bsym{\beta}^p$. In other words, by inserting dislocations in the originally defect-free medium we
are not allowed to create microcracks that would destroy the continuity of the body. This
requirement is equivalent to demanding that the total distortion field $\bsym{\beta}^t$ be
curl-free, i.e.
\begin{equation}
  \nabla \times \bsym{\beta}^t = \bsym{0} \quad.
  \label{eq_curl_betat}
\end{equation}
Here, $\bsym{\beta}^t =
\bsym{\beta} + \bsym{\beta}^p$ is the total plastic distortion\footnote{In the following, we will
  show that both elastic and plastic parts of the distortion tensor contain signatures of individual
  dislocations. To simplify the notation, we avoid the superscript $e$ that often labels the elastic
  part of the distortion tensor.}  that is written as a sum of its elastic part $\bsym{\beta}$ and
the plastic part $\bsym{\beta}^p$.  Each of these distortions can be expressed as a sum of their
symmetric parts, i.e. strains $\bsym{\eps}^t$, $\bsym{\eps}$, $\bsym{\eps}^p$, and antisymmetric
parts corresponding to rotations $\bsym{\omega}^t$, $\bsym{\omega}$, $\bsym{\omega}^p$.

\begin{figure}[!htb]
  \centering
  \includegraphics[scale=.9]{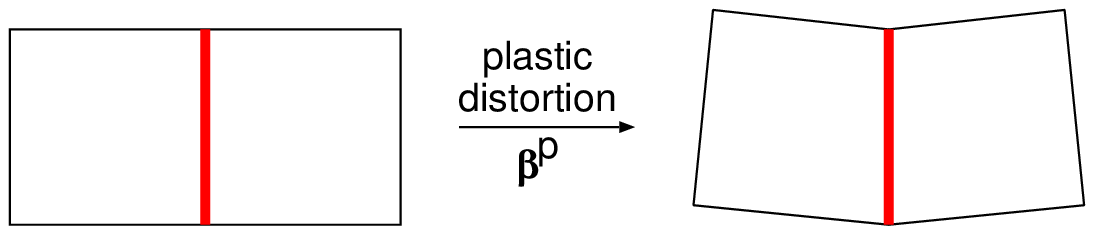} \\
  (a) compatible plastic distortion \\[1em]
  \includegraphics[scale=.9]{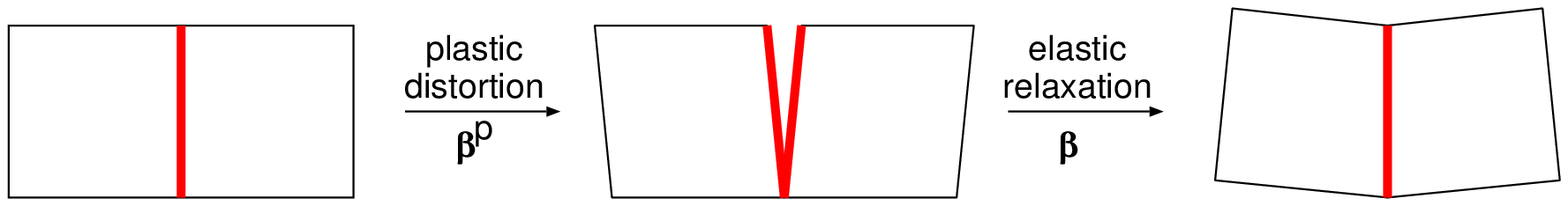} \\
  (b) incompatible plastic distortion
  \caption{Two kinds of plastic distortion of two neighboring mesoscopic cells. The compatible
    plastic distortion (a) maintains the continuity of the body, while the incompatible plastic
    distortion (b) requires elastic relaxation to keep the body crack-free.}
  \label{fig_distort}
\end{figure}

In general, two kinds of plastic distortions of a mesoscopic cell can take place depending on the
way the adjacent cells are distorted and these are shown in \reffig{fig_distort}. If the net Burgers
vector in a given cell is the same as those in the adjacent cells, all these cells are distorted the
same way and, therefore, the continuity of the body is locally preserved.  Hence, $\nabla \times
\bsym{\beta^p} = \bsym{0}$ and, with the help of \refeq{eq_curl_betat} and $\bsym{\beta}^t =
\bsym{\beta} + \bsym{\beta}^p$, this means that the elastic part of the strain field is also
curl-free, i.e. $\nabla\times\bsym{\beta} = \bsym{0}$. In this case, the above-mentioned requirement
of the continuous medium leads to the well-known Saint-Venant elastic compatibility
constraint\footnote{Tensorial representations of divergence, curl and incompatibility are given in
  Appendix~\ref{apx_tensors}.}
\begin{equation}
  \nabla \times \nabla \times \bsym{\eps} = \bsym{0} \quad.
  \label{eq_eps_comp}
\end{equation}
Since the elastic strains are compatible in the sense of the Saint-Venant law, this plastic
distortion is referred to as \emph{compatible}\footnote{The Saint-Venant law is satisfied for all
  irrotational strain fields, for example those due to point defects.}. In contrast, one can imagine
a more general case where the adjacent cells are characterized by different net Burgers vectors,
which means that the cells are distorted differently. Hence, the plastic part of the distortion
tensor is no longer curl-free and, instead, $-\nabla \times \bsym{\beta}^p = \bsym{\alpha}$, where
$\bsym{\alpha}$ is the tensor of the density of net Burgers vectors\footnote{There seem to be two
  equivalent definitions of this tensor in the literature. In the original Kr\"oner's formulation,
  $\bsym{\alpha} = -\nabla \times \bsym{\beta}$ and the incompatibility of strains was then defined
  as $\bsym{\eta} = \nabla \times \bsym{\eps} \times \nabla$. Here, we adopt the convention of
  \citet{el-azab:00} whereby $\bsym{\alpha} = \nabla \times \bsym{\beta}$, and this allows us to
  write the incompatibility constraint in a more intuitive way as $\bsym{\eta} = \nabla \times
  \nabla \times \bsym{\eps}$. }.  If this plastic distortion acted alone, it would cause disregistry
between neighboring mesoscopic cells\cite{kroner:58} and thus contradiction of the requirement that
we set forth by \refeq{eq_curl_betat}. However, cohesive forces of the matter act to remove this
disregistry and this relaxation proceeds purely elastically.  Clearly, in order to satisfy
\refeq{eq_curl_betat}, the elastic strain field cannot be arbitrary but has to satisfy the
constraint $\nabla \times \bsym{\beta} = \bsym{\alpha}$. Performing the curl of this equation and
taking its symmetric part then leads to an incompatibility constraint between the components of the
elastic strain tensor,
\begin{equation}
  \nabla \times \nabla \times \bsym{\eps} = \bsym{\eta} \quad,
  \label{eq_eps_incomp}
\end{equation}
where $\bsym{\eta}$ is the so-called incompatibility tensor defined as
\begin{equation}
  \bsym{\eta} = \sym (\nabla \times \bsym{\alpha}) \quad.
\label{eq_etadef}
\end{equation}
In this case, the individual components of the elastic part of the strain tensor, $\bsym{\eps}$, are
not compatible in the sense of the Saint-Venant law \refeq{eq_eps_comp}. Hence, this plastic
distortion is called \emph{incompatible} and the degree of this incompatibility is quantified by
the symmetric tensor $\bsym{\eta}$.

It is important to emphasize that since each net Burgers vector is a sum of many Burgers vectors of
the underlying crystal dislocations, this coarse-graining procedure essentially determines the
vector corresponding to the excess of the crystallographic Burgers vectors. Moreover, since each
mesoscopic cell contains a large number of crystal dislocations, the density of net Burgers vectors
can be approximated by a tensor field that is continuous throughout the entire space. In the
Cartesian coordinate system with axes $x_1$, $x_2$, $x_3$, this density is represented by the
Nye\cite{nye:53} tensor $\alpha_{ij}$, the components of which relate to the net dislocations with
line directions parallel to the $x_i$ axis and the Burgers vectors parallel to the $x_j$ axis.
Hence, the diagonal elements of $\bsym{\alpha}$ correspond to screw components, while the
off-diagonal elements to edge components of the net Burgers vectors. One can thus determine the
density of the net Burgers vectors as $\alpha_{ij} = B_j / S_i$, where $B_j$ is the $j$-th component
of the net Burgers vector and $S_i$ are the components of the vector normal to the oriented area of
the mesoscopic cell pinned by the dislocation line. The tensor $\alpha_{ij}$ should not be confused
with the dislocation density $\rho$ that is, by definition, the total length of all dislocation
lines that populate the medium divided by its volume.

For simplicity, consider now a two-dimensional plane strain problem in which the only nonzero
components of the elastic strain tensor are $\eps_{11}$, $\eps_{12}$, and $\eps_{22}$. Hence, it is
straightforward to prove that the only component equation of \refeq{eq_eps_incomp} that is not
identically zero is
\begin{equation}
  \partial_{22}\eps_{11} - 2\partial_{12}\eps_{12} + \partial_{11}\eps_{22} = \eta_{33} \quad ,
  \label{eq_incompat}
\end{equation}
where $\partial_{ij} \equiv \partial^2/\partial{x_i}\partial{x_j}$. It is important to recognize
that the scalar incompatibility field $\eta_{33}$ is nonzero wherever the distribution of 
dislocations in the medium causes the plastic distortion to be incompatible in the sense explained
above. Hence, the relation \refeq{eq_incompat} represents a constraint that the elastic strains have
to satisfy in order to maintain the continuity of the body that has been broken locally by the
incompatible plastic distortion. From \refeq{eq_etadef}, the incompatibility field $\eta_{33}$ can
be determined as
\begin{equation}
  \eta_{33} = \partial_1\alpha_{32} - \partial_2\alpha_{31} \quad ,
  \label{eq_eta33}
\end{equation}
where $\partial_i \equiv \partial/\partial{x_i}$. Since only the components $\alpha_{31}$ and
$\alpha_{32}$ appear in this expression, only edge dislocations with their line directions parallel
to $x_3$ and the Burgers vector components along the $x_1$ and $x_2$ axes contribute to the elastic
strain incompatibility in this two-dimensional case.

In order to develop a clear link between the microscopic crystal dislocations and their
coarse-grained mesoscopic manifestation by the net Burgers vector $\bsym{B}$, we will now consider a
finite number of discrete slip systems $s$. In each such system, the Burgers vector of crystal
dislocations $\bsym{b}^s$ is known, e.g. $\bsym{b}^s=1/2\langle{110}\rangle$ (in units of the
lattice parameter $a_0$) for edge dislocations in face-centered cubic crystals. To each mesoscopic cell we
can then attribute the net Burgers vector
\begin{equation}
  \bsym{B}(\bsym{r}) = \sum_s N^s(\bsym{r}) \bsym{b}^s \quad, 
  \label{eq_B_slipsys}
\end{equation}
where $N^s$ is the number of crystal dislocations in the slip system $s$ with the Burgers vector
$\bsym{b}^s$. Instead of working with integral values $N^s$, it is convenient to define a number
density of crystal dislocations in the slip system $s$ as $n^s=N^s/S_{cell}$, where $S_{cell}$ is
the area of a mesoscopic cell, and regard this as a continuous variable. Hence, the density of net
Burgers vectors can be written as
\begin{equation}
  \alpha_{3i}(\bsym{r}) = B_i(\bsym{r})/S_{cell} = \sum_s n^s(\bsym{r}) b_i^s \quad.
  \label{eq_alpha_slipsys}
\end{equation}
Substituting this expression in \refeq{eq_eta33} yields the strain incompatibility in terms of the
density of crystal dislocations in individual slip systems:
\begin{equation}
  \eta_{33}(\bsym{r}) = \epsilon_{ij} \sum_s \frac{\partial n^s(\bsym{r})}{\partial x_i} b_j^s \quad,
  \label{eq_eta33_slipsys}
\end{equation}
where $\epsilon_{ij}$ is the Levi-Civita tensor. The expression \refeq{eq_eta33_slipsys} represents
a ``recipe" for coarse-graining the density of crystal dislocations $n^s$ in individual slip systems $s$
into a continuously distributed incompatibility field $\eta_{33}$. It is important to emphasize that
each slip system $s$ contains crystal dislocations with positive and negative Burgers vectors of the
same magnitude. The dislocation density $n^s$ can thus be written as $n^s = n^{s+}-n^{s-}$, where
$n^{s+}$ and $n^{s-}$ are non-negative densities of crystal dislocations with positive
($\bsym{b}^s$) and negative ($-\bsym{b}^s$) Burgers vectors, respectively. In
Section~\ref{sec_relax_alpha} we will show that this distinction between positive and negative
Burgers vectors of crystal dislocations is required for a systematic evolution of the dislocation
density.


\section{Mesoscopic free energy}
\label{sec_freeE}

Consider an elastically anisotropic body of which every element is subjected to a generally nonuniform
stress tensor $\bsym{\sigma}$ and the response to this particular loading is characterized in each
such element by the elastic strain tensor $\bsym{\eps}$. If we consider that a linear relation
between the applied stress and induced strain applies, the free energy of this deformed medium
equals the total strain energy, i.e.  $F = \int_V \frac{1}{2}\sigma_{ij}\eps_{ij} \d{\bsym{r}}$.
Writing $\sigma_{ij} = c_{ijkl}\eps_{kl}$, where $c_{ijkl}$ is the elastic stiffness tensor, one
arrives at the free energy \cite{landau:86}
\begin{equation}
  F = \int_V \frac{1}{2} c_{ijkl} \eps_{ij} \eps_{kl} \d{\bsym{r}} \quad.
  \label{eq_freeE_harm}
\end{equation}
The elastic stiffness tensor has generally 21 independent elastic constants but any symmetry of the
underlying crystal structure reduces this number. If the stress tensor is written in the Voigt
notation as $\sigma_i = [\sigma_{11}\ \sigma_{22}\ \sigma_{33}\ \sigma_{23}\ \sigma_{13}\
\sigma_{12}]^T$ and the strain tensor as $\eps_j = [\eps_{11}\ \eps_{22}\ \eps_{33}\ \eps_{23}\
\eps_{13}\ \eps_{12}]^T$, the free energy \refeq{eq_freeE_harm} can be expressed equivalently as $F
= \int_V \frac{1}{2} C_{ij} \eps_i \eps_j \d{\bsym{r}}$, where $C_{ij}$ is the $(6 \times 6)$
symmetric elastic stiffness matrix. For simplicity, we will be concerned in the following with cubic
symmetry and this is characterized by three independent elastic constants $C_{11}$, $C_{12}$, and
$C_{44}$. The corresponding free energy for cubic symmetry then reads
\begin{equation}
  F = \int_V \left\{ \frac{1}{2} C_{11} (\eps_{11}^2 + \eps_{22}^2) + C_{12} \eps_{11} \eps_{22} + 
    2C_{44} \eps_{12}^2 \right\} \d{\bsym{r}} \quad ,
  \label{eq_freeE_strains}
\end{equation}
where we write the strains again in their usual two-index notation.

\begin{figure}[!htb]
  \begin{minipage}{.3\textwidth}
    \includegraphics{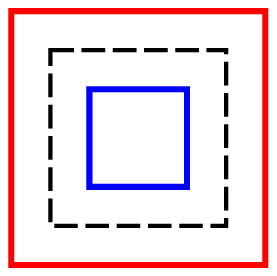} \\
    $e_1$
  \end{minipage}
  \begin{minipage}{.3\textwidth}
    \includegraphics{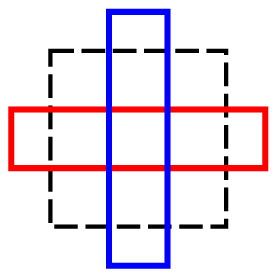} \\
    $e_2$
  \end{minipage}
  \begin{minipage}{.3\textwidth}
    \includegraphics{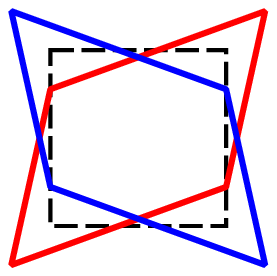} \\
    $e_3$
  \end{minipage}
  \caption{The three order parameters used to study the square to rectangle phase transformation in
    crystals of cubic symmetry; {\color{red}red}=positive, {\color{blue}blue}=negative value.}
  \label{fig_e1e2e3}
\end{figure}

For simplicity, we will specialize in the following to the case of the square to rectangle phase
transformations\footnote{A similar approach is valid also in higher dimensions, where typically more
  than one primary order parameter is needed to identify the phase transition.} that can be thought
of as a two-dimensional reduction of the cubic to tetragonal (or tetragonal to orthorhombic) phase
transformation frequently observed in shape memory alloys such as Fe-Pd, In-Tl, Ni-Al(Mn), ternary
alloys Fe-Ni-C, magnetoelastic alloy Ni$_2$MnGa, or even perovskites such as BaTiO$_3$ that exhibit
strain-induced polarization. A naive way to identify a phase transformation would be to define the
order parameter as a ratio of two lattice parameters. However, this ratio alone would not
distinguish between cubic and rhombohedral symmetry and, therefore, one order parameter is generally
not sufficient to identify the crystal structure.  In the following, we consider three scalar order
parameter fields\cite{kartha:95},
\begin{equation}
  e_1 = \frac{1}{\sqrt{2}}(\eps_{11}+\eps_{22}) \quad,\quad
  e_2 = \frac{1}{\sqrt{2}}(\eps_{11}-\eps_{22}) \quad,\quad
  e_3 = \eps_{12} \quad,
  \label{eq_op}
\end{equation}
where $e_1$ measures the isotropic dilation, $e_2$ the deviatoric change of shape, and $e_3$ the
change of the right angle caused by the shear. These three fundamental modes of deformation of an
element in the body are shown in \reffig{fig_e1e2e3}, where $e_2=0$ corresponds to the austenite and
negative/positive $e_2$ to the two variants of the martensite.  Here, $e_2$ serves as the primary
order parameter for the square to rectangle transformation, whereas $e_1$ and $e_3$ are secondary
order parameters. In terms of these fields, the Hookean elastic free energy \refeq{eq_freeE_strains}
for this two-dimensional problem is
\begin{equation}
  F = \int_S \left\{ \frac{A_1}{2}e_1^2 + \frac{A_2}{2}e_2^2 + \frac{A_3}{2}e_3^2 \right\} 
  \d{\bsym{r}} \quad,
  \label{eq_freeE_harm_final}
\end{equation}
where the coefficients are related to the elastic constants\footnote{In the isotropic case,
$C_{11}-C_{12}=2C_{44}$ and, therefore, only two elastic constants are independent. Hence, 
only two of the three coefficients $A_1$, $A_2$, $A_3$ in the free energy are independent since 
$A_2=A_3/2$.} as $A_1 = C_{11}+C_{12}$, $A_2 = C_{11}-C_{12}$, and $A_3 = 4C_{44}$.

For studies of first order phase transitions the order parameter $e_2$ is expanded in even powers up
to the sixth order as odd powers are not allowed by symmetry. In addition, it is customary
\cite{kartha:95,shenoy:99,rasmussen:01,lookman:03} to incorporate a gradient term proportional to
$|\nabla e_2|^2$ that represents the energy cost for spatial variation of the order parameter, and
the strain energy due to coupling of the internal strain with the externally applied stress
field. Consequently, the free energy can be written as
\begin{equation}
  F = \int_S \left\{ f_{loc}[e_2] + f_{nonloc}[e_1,e_3] + f_{grad}[e_2] - f_{load}[e_1,e_2,e_3] \right\} 
  \d{\bsym{r}} \quad,
  \label{eq_freeE}
\end{equation}
where the various energy densities are
\begin{eqnarray}
  \label{eq_fdensity}
  \nonumber
  f_{loc}[e_2] &=& \frac{A_2}{2} e_2^2 + \frac{B}{4} e_2^4 + \frac{C}{6} e_2^6 \quad,\\
  \nonumber
  f_{nonloc}[e_1,e_3] &=& \frac{A_1}{2} e_1^2 + \frac{A_3}{2} e_3^2 \quad,\\
  f_{grad}[e_2] &=& \frac{K_2}{2} |\nabla e_2|^2 \quad,\\
  \nonumber
  f_{load}[e_1,e_2,e_3] &=& \frac{1}{2}\sigma_{ij}\eps_{ij} \quad.
\end{eqnarray}
It is important to emphasize that the strain energy density as given by $f_{load}$ applies only to
the linear-elastic case, i.e. when the internal strains induced in the body by the external loading
vary linearly with this applied stress.

In defect-free media the individual components of the elastic strain tensor $\bsym{\eps}$ are
related by the Saint-Venant constraint \refeq{eq_eps_comp} and this guarantees that the strains can
be determined from the known displacement field by taking its gradients. In contrast, the presence
of dislocations causes discontinuities in the displacement field and, therefore, the integrability
condition no longer applies.  Hence, the strains are incompatible in the sense of the Saint-Venant
condition and this is expressed by \refeq{eq_eps_incomp}, where $\bsym{\eta}$ represents the
``strength" of this incompatibility. In the case of plane strain in the $(x_1,x_2)$ plane the
incompatibility constraint \refeq{eq_incompat} can be expressed using \refeq{eq_op} in terms of the
order parameters as
\begin{equation}
  \nabla^2 e_1 - (\partial_{11}-\partial_{22})e_2 - \sqrt{8} \partial_{12} e_3 = \eta_{33}\sqrt{2} \quad.
  \label{eq_eincomp}
\end{equation}
Now, suppose that we know the density of crystal dislocations $n^s$ in each slip system $s$. Hence,
the corresponding incompatibility field $\eta_{33}$, obtained from \refeq{eq_eta33_slipsys}, is
known at every point.  We then seek the fields $e_1$, $e_2$, $e_3$ that minimize the free energy
\refeq{eq_freeE} subject to the incompatibility constraint \refeq{eq_eincomp}. The additive nature
of the free energy allows us to perform this minimization in steps. Firstly, we calculate the fields
$e_1$ and $e_3$ that are constrained by the incompatibility condition \refeq{eq_eincomp} and, in the
second step, we obtain the primary order parameter $e_2$ by minimizing the free energy. The part of
the free energy that depends explicitly on $e_1$ and $e_3$ reads
\begin{equation}
  F_{13} = \int_S \left\{ f_{nonloc}[e_1,e_3] + f_{load}[e_1,e_2,e_3] + \lambda G[e_1,e_2,e_3,\eta_{33}] 
  \right\} \d{\bsym{r}} \quad,
\end{equation}
where $\lambda$ is the Lagrange multiplier that incorporates the incompatibility constraint $G =
\nabla^2 e_1 - (\partial_{11}-\partial_{22})e_2 - \sqrt{8} \partial_{12} e_3 -
\eta_{33}\sqrt{2} = 0$. The fields $e_1$ and $e_3$ that minimize $F_{13}$ are then obtained from the
stationary conditions\footnote{If we consider periodic boundary conditions, the integration by parts
  transfers the derivatives of the variations of $e_1$ and $e_3$ to the derivatives of
  $\lambda$, e.g. $\lambda \nabla^2 \delta e_1$ becomes $(\nabla^2 \lambda) \delta e_1$, etc.}
$\delta{F_{13}} / \delta{e_1} = 0$, $\delta{F_{13}} / \delta{e_3} = 0$, and $\delta{F_{13}} /
\delta{\lambda} = 0$. When evaluated in k-space, these conditions provide closed-form expressions
for the secondary order parameter fields $e_1$, $e_3$ in terms of the incompatibility $\eta_{33}$
and the components of the externally applied stress tensor $\sigma_{ij}$:
\begin{equation}
  e_i(\k) = Q_i(\k) e_2(\k) - \sqrt{2} R_i(\k) \eta_{33}(\k) -
  S_i(\k)\sigma_{jj}(\k) - T_i(\k)\sigma_{12}(\k)\quad,
  \label{eq_eik}
\end{equation}
where $i=1,3$. Here, $Q_1 \sim k_x^4-k_y^4$, $Q_3 \sim k_xk_y(k_x^2-k_y^2)$, $R_1 \sim k^2$, $R_3
\sim k_xk_y$, $S_1 \sim k^4$, $S_3 \sim k_xk_yk^2$, $T_1 \sim k_xk_yk^2$, and $T_3 \sim k_x^2k_y^2$
are k-space kernels that we write explicitly in \refapx{apx_kernels}, and $\sigma_{jj} = \sigma_{11}
+ \sigma_{22}$. Eq.  \refeq{eq_eik} represents the most general form\footnote{If the incompatibility
  field $\eta_{33}$ vanishes, i.e.  the medium is dislocation-free, and no external stress is
  applied, only the first term in \refeq{eq_eik} remains. In this case, one obtains\cite{kartha:95}
  that $e_1$ and $e_3$ can be expressed as functionals of the primary order parameter $e_2$ only.}
valid for the plane strain case in which the stress tensor has all components nonzero.  With the
help of \refeq{eq_eik}, the nonlocal part of the free energy in \refeq{eq_fdensity} can be written
as a functional of the primary order parameter $e_2$, the incompatibility $\eta_{33}$, and the
components of the stress tensor $\sigma_{ij}$ only:
\begin{eqnarray}
\nonumber
  f_{nonloc}(\k) = \frac{A_{13}(\k)}{2} [e_2(\k)]^2 - 
  \sqrt{2} B_{13}(\k) e_2(\k) \eta_{33}(\k) - \Sigma_{AQ}(\k) e_2(\k) + \\
  + C_{13}(\k) [\eta_{33}(\k)]^2 + \sqrt{2}\Sigma_{AR}(\k)\eta_{33}(\k) + 
  \frac{\Sigma_A(\k)}{2} \quad,
  \label{eq_fnonloc_final}
\end{eqnarray}
where the k-space kernels $\Sigma_{AQ}$, $\Sigma_{AR}$ and $\Sigma_A$, written explicitly in
\refapx{apx_kernels}, depend on the external stress field. In \refeq{eq_fnonloc_final}, the first
term represents nonlocal interactions in the $e_2$ field, whereas the second and third terms are
couplings of $e_2$ with the incompatibility $\eta_{33}$ and the external stress $\sigma_{ij}$,
respectively. The fourth and the fifth terms are contributions from the incompatibility and its
coupling to the external stress field, and the last term represents the shift of the free energy by
the external stress field. If the incompatibility $\eta_{33}$ vanishes and no external stress is
applied, only the first term remains and we recover the nonlocal expression for an unloaded defect-free
medium\cite{kartha:95, lookman:03}.

In a similar way as we expressed $f_{nonloc}$ in terms of $e_2$, $\eta_{33}$, and $\sigma_{ij}$
only, we can utilize \refeq{eq_eik} to obtain a reduced expression for the strain energy density
$f_{load}$. For plane strain, a completely general stress state leads to $f_{load} =
\frac{1}{2}\sigma_{11}\eps_{11} + \sigma_{12}\eps_{12} + \frac{1}{2}\sigma_{22}\eps_{22}$.
Expressing the strains in terms of the order parameters $e_1$, $e_2$, $e_3$, and using
\refeq{eq_eik}, one arrives at the strain energy density
\begin{equation}
  f_{load}(\k) =  W_Q(\k) e_2(\k) - \sqrt{2} W_R(\k) \eta_{33}(\k) - W_\Sigma(\k) \quad,
  \label{eq_fload_final}
\end{equation}
where the kernels $W_Q$, $W_R$ and $W_\Sigma$ are again written explicitly in \refapx{apx_kernels}.
The remaining two free energy densities, i.e. $f_{loc}$ and $f_{grad}$, are functionals of $e_2$
only and are determined uniquely by \refeq{eq_fdensity}.


\section{Relaxation of the primary order parameter field}
\label{sec_relax_e2}

\begin{figure}[!htb]
  \centering
  \includegraphics[width=10cm]{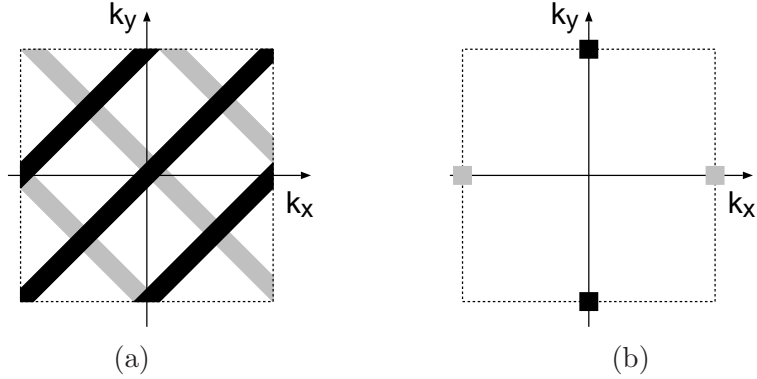} \\
  (a) \hskip6cm (b)
  \caption{Schematic illustration of the texture in $e_2$ that minimizes the first two terms in the
    nonlocal part of the free energy density \refeq{eq_fnonloc_final}. (a) minimizes
    $A_{13}(\k)[e_2(\k)]^2$ -- the two orientations of twins yield the same free energy; (b)
    minimizes $-B_{13}(\k)e_2(\k)\eta_{33}(\k)$ -- gray squares correspond to
    $e_2(\k)\eta_{33}(\k) > 0$ and black squares to $e_2(\k)\eta_{33}(\k) < 0$.}
  \label{fig_frustr}
\end{figure}

Since all constituents of the free energy are now functionals of the primary order parameter $e_2$,
the incompatibility $\eta_{33}$ and the stress tensor $\sigma_{ij}$ and, assuming that the
incompatibility and the external stress field change slowly relative to $e_2$, i.e. they remain
approximately constant on the time scale of relaxation of $e_2$, it is straightforward to find the
field $e_2$ that minimizes the free energy.  This minimization, i.e. the solution of the equations
$\delta F/\delta e_2 = 0$ and $\delta^2 F/\delta e_2^2 > 0$, cannot be performed
analytically. However, we may formulate a relaxational dynamics\footnote{It is always possible to
  augment this deterministic relaxational dynamics by a stochastic term that is often
  characterized\cite{sagues:07} as white noise with zero mean and variance $2\Gamma k_B
  T$. However, it can be shown that this noise plays an important role only at the temperatures
  slightly below $T_c$, where thermal fluctuations may overcome the energy barrier between the two
  variants of the martensite and, therefore, cause switching between these variants. In this case,
  the twin boundary between individual martensite variants would not be sharp but rather diffuse. At
  low temperatures, i.e. well below $T_c$, this barrier is large and the weak thermal fluctuations
  cannot cause this switching.  Similarly, above $T_c$, the free energy has one minimum
  corresponding to $e_2=0$ and the thermal fluctuations would merely cause broadening of the
  distribution of $e_2$. The goal of this paper is to give a proof of the principle and, for
  simplicity, the effect of thermal noise is not included.} for $e_2$ that will follow the path of
the steepest descent of the free energy $F$:
\begin{equation}
  \frac{\partial e_2}{\partial t} = -\Gamma \frac{\delta F}{\delta e_2} \quad,
  \label{eq_e2evolve}
\end{equation}
where $\Gamma$ plays a role of the mobility parameter. Writing \refeq{eq_e2evolve} as a difference
scheme, i.e. $e_2(t+\Delta t) = e_2(t) - \Delta t \Gamma\, \delta F/\delta e_2$, it is clear that
$\Gamma$ merely renormalizes the time step $\Delta t$. Moreover, since $\delta F/\delta e_2 =
\partial(f_{loc} + f_{nonloc} + f_{grad} - f_{load})/\partial e_2$, the right-hand side of
\refeq{eq_e2evolve} can be calculated easily by taking derivatives of the previously derived free
energy densities:
\begin{eqnarray}
  \nonumber
  \frac{\partial f_{loc}(\rr)}{\partial e_2(\rr)} &=& A_2 e_2(\rr) + B [e_2(\rr)]^3 + C [e_2(\rr)]^5 \quad,\\
  \nonumber
  \frac{\partial f_{nonloc}(\k)}{\partial e_2(\k)} &=& A_{13}(\k) e_2(\k) - \sqrt{2} B_{13}(\k) \eta_{33}(\k) - 
    \Sigma_{AQ}(\k) \quad,\\
  \frac{\partial f_{grad}(\rr)}{\partial e_2(\rr)} &=& -K_2 \nabla^2 e_2(\rr) \quad,\\
  \nonumber
  \frac{\partial f_{load}(\k)}{\partial e_2(\k)} &=& W_Q(\k) \quad.
\end{eqnarray}
If one considers a defect-free medium and no external stress is applied, the second and the third
terms on the right-hand side of $\partial f_{nonloc}/\partial e_2$ vanish and also $\partial
f_{load}/\partial e_2=0$. In this case, the nonlocal free energy density \refeq{eq_fnonloc_final}
reduces to $f_{nonloc} = (A_{13}(\k)/2) [e_2(\k)]^2$ which is identical to the form obtained by
\citet{kartha:95} Recognizing that the kernel $A_{13}(\k) \sim (k_x^2-k_y^2)^2$ is minimized when
$k_x = \pm k_y$, it directly follows that the system minimizes its free energy by aligning the
nonzero components of the order parameter field $e_2$ along any of the two k-space diagonals, as
shown in \reffig{fig_frustr}a.  The system thus develops diagonal striations in real space, similar
to the tweed microstructure that is a general feature of many martensites\cite{gushchin:84,
  greenberg:03, xu:04, sehitoglu:03, daly:07, inamura:07, field:01, gall:98}.

The presence of dislocations changes the evolving order parameter texture considerably and this can
be demonstrated by examining the first three terms in \refeq{eq_fnonloc_final}. As mentioned above,
the first term is minimized when all nonzero components of $e_2(\k)$ are aligned along the k-space
diagonal, as shown in \reffig{fig_frustr}a. In the second term, the kernel $-B_{13}(\k) \sim
-k_x^2+k_y^2$ is minimized when $\bsym{k}= (\pm \pi/l,0)$ and maximized when $\bsym{k} =
(0,\pm\pi/l)$, where $l$ is the real-space width of the simulated domain. Therefore, the second term
in \refeq{eq_fnonloc_final} is minimized when positive $e_2(\k)\eta_{33}(\k)$ is at the minimum of
$-B_{13}(\k)$, i.e. at $\bsym{k} = (\pm\pi/l,0)$ and, simultaneously, negative
$e_2(\k)\eta_{33}(\k)$ is at its maximum, i.e. at $\bsym{k} = (0,\pm\pi/l)$. This is shown in
\reffig{fig_frustr}b. In the real space the minimum of the second term in \refeq{eq_fnonloc_final}
thus corresponds to a three-state ``checkerboard'' pattern with the periodicities along the $x_1$ and $x_2$ axes
equal to twice the width of the mesoscopic cell. The size of the mesoscopic cell thus imposes an
intrinsic length scale in the order parameter field. Similarly, one can identify the pattern that
minimizes the third term in \refeq{eq_fnonloc_final} which now depends on the external stress
field. Since the first three terms in $f_{nonloc}$ cannot be minimized simultaneously when a finite
incompatibility is introduced, they naturally compete with each other.  Hence, the minimization of
the free energy by the field $e_2$ subject to a fixed distribution of dislocations is inherently
frustrated and $e_2$ does not always evolve into a well-defined diagonal texture as it does in
defect-free materials. We will see later that the diagonal texture is preferred at low dislocation
densities, whereas at high dislocation densities the terms containing the incompatibility
$\eta_{33}$ become significant and the texture in $e_2$ tends to that corresponding to
\reffig{fig_frustr}b.


\section{Evolution of the dislocation density}
\label{sec_relax_alpha}

\begin{figure}[!htb]
  \centering
  \includegraphics[width=5cm]{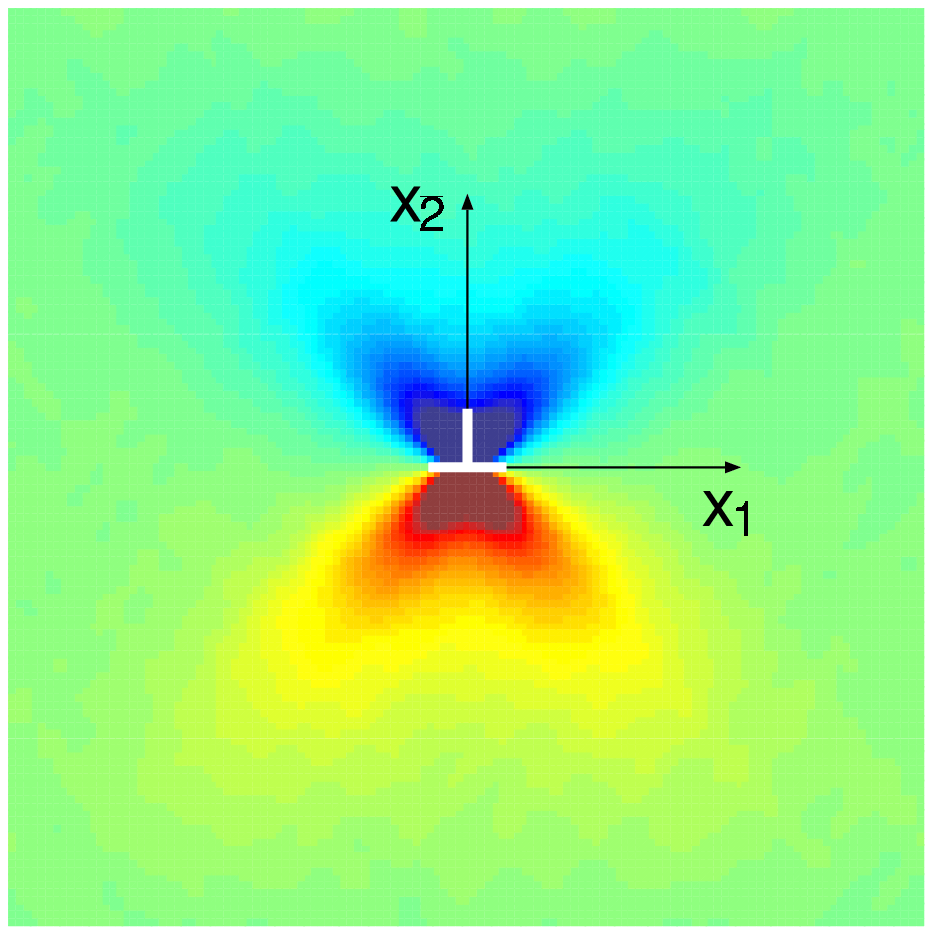} \quad
  \includegraphics[width=5cm]{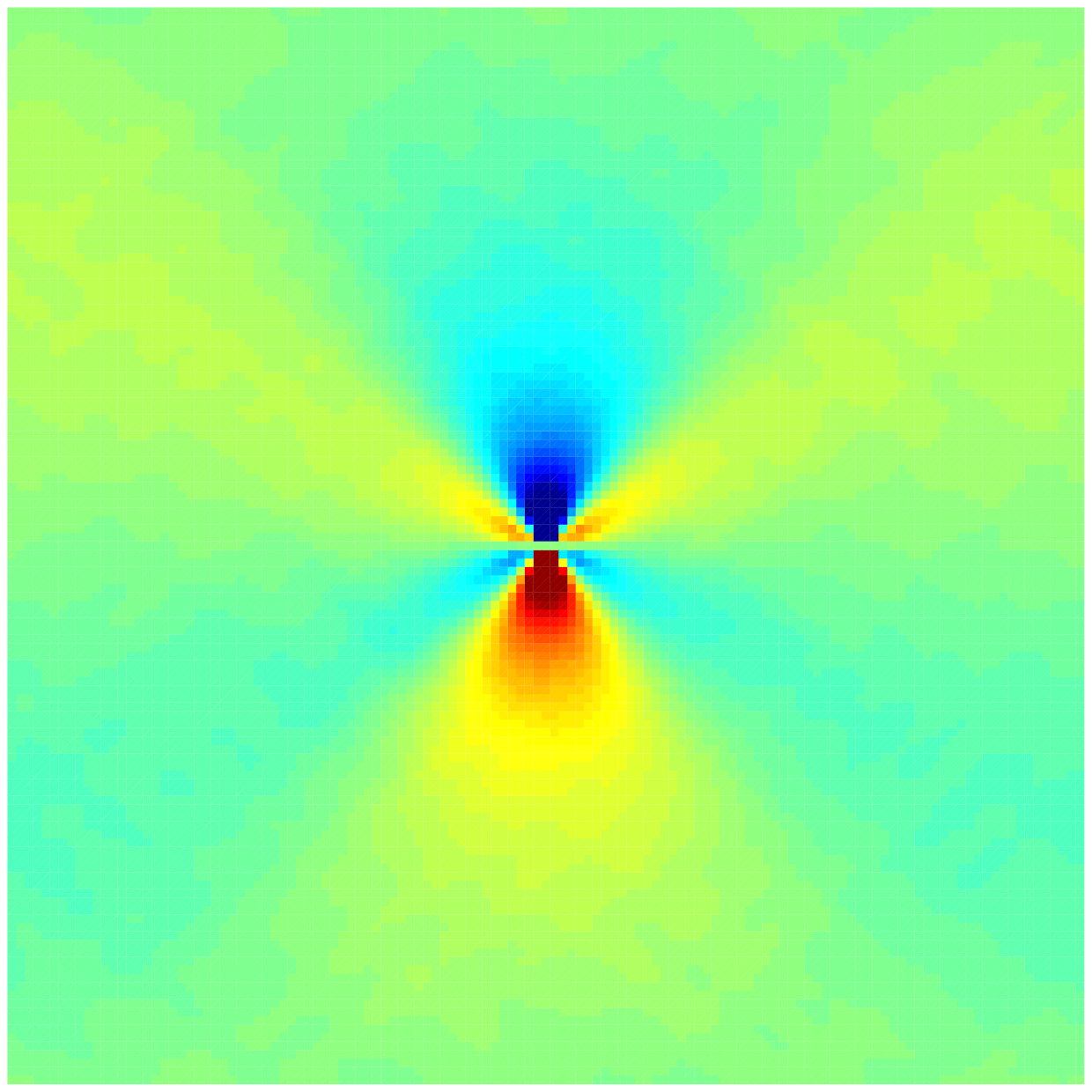} \quad
  \includegraphics[width=5cm]{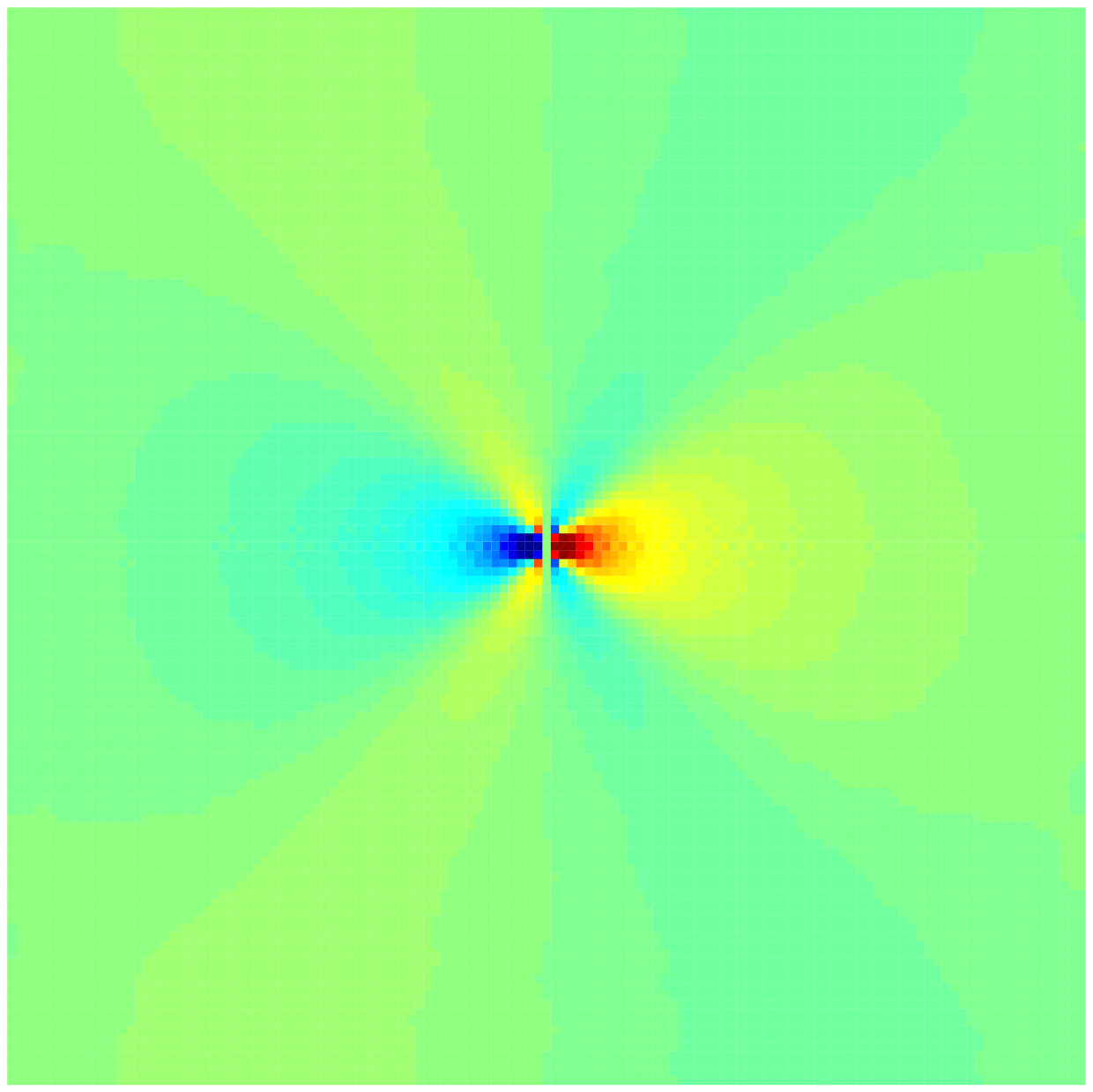} \\
  (a) $\sigma_{11}(\rr)$ \hskip4cm (b) $\sigma_{22}(\rr)$ \hskip4cm (c) $\sigma_{12}(\rr)$
  \caption{Stress field around one edge dislocation with its Burgers vector parallel to the $x_1$
    axis ({\color{blue}blue}=negative, {\color{red}red}=positive values).  Because periodic boundary
    conditions are used in this calculation, the stress field shown here corresponds to one
    dislocation from a periodic array of dislocations of the same kind.}
  \label{fig_sigedge}
\end{figure}

From the order parameter field $e_2$ that minimizes the free energy, we can obtain $e_1$ and $e_3$
using \refeq{eq_eik} and subsequently $\eps_{11}$, $\eps_{12}$ and $\eps_{22}$ using
\refeq{eq_op}. Assuming linear-elastic dependence between the stresses and strains and using the
three anisotropic elastic constants $C_{11}$, $C_{12}$ and $C_{44}$, the components of the internal
stress field are obtained from:
\begin{eqnarray}
  \label{eq_sigint}
  \left[
  \begin{array}{c}
    \sigma_{11}(\rr) \\
    \sigma_{22}(\rr) \\
    \sigma_{12}(\rr)
  \end{array}
  \right]
  &=&
  \left[
  \begin{array}{ccc}
    C_{11} & C_{12} & 0 \\
    C_{12} & C_{11} & 0 \\
    0 & 0 & C_{44}
  \end{array}
  \right]
  \left[
  \begin{array}{c}
    \eps_{11}(\rr) \\
    \eps_{22}(\rr) \\
    \eps_{12}(\rr)
  \end{array} 
  \right] \quad, \\
  \nonumber
  \sigma_{33}(\rr) &=& \frac{C_{12}}{C_{11}+C_{12}} [\sigma_{11}(\rr)+\sigma_{22}(\rr)] \quad.
\end{eqnarray}
Note that just by minimizing the free energy subject to the elastic strain incompatibility
constraint we obtained an internal stress field that is a superposition of the elastic stress fields
of individual net dislocations. In order to demonstrate this, we show in \reffig{fig_sigedge} the
calculated stress field around one edge dislocation in an isotropic medium with its Burgers vector
along the positive $x_1$ direction.  The overall distribution of stresses is in excellent agreement
with the formulas derived from isotropic elasticity (see \citet{hirth:82}). However, in the
derivation of the free energy no constraints were imposed on the anisotropy of elastic constants
and, therefore, the stress fields of the dislocations in arbitrarily anisotropic media can also be
calculated just by minimizing the free energy.

\begin{figure}[!htb]
  \centering
  \includegraphics[scale=.8]{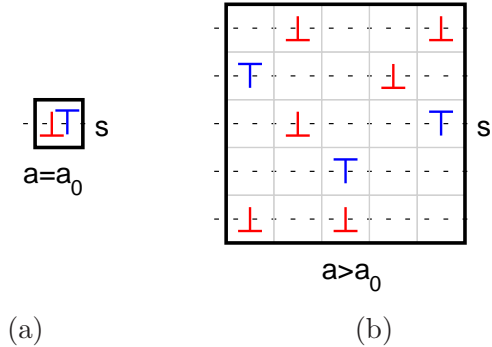} \\
  \hskip-1.5cm (a) \hskip4cm (b)
  \caption{Schematic illustration of the slip planes (dotted lines) corresponding to a particular
    slip system $s$ passing through the mesoscopic cell (black square) when it comprises only one
    (a) and many (b) crystallographic unit cells (gray squares).}
  \label{fig_mesocell_size}
\end{figure}

It is important to realize that each slip system contains crystal dislocations with positive and
negative Burgers vectors and in our mesoscopic description we have to treat the densities of these
dislocations separately. The explanation is provided in \reffig{fig_mesocell_size} where we show
schematically two different sizes of the mesoscopic cell and the corresponding parallel slip planes
(dotted lines) that pass through these cells. If the mesoscopic cell is of the same size as the
crystallographic unit cell (\reffig{fig_mesocell_size}a), only one slip plane corresponding to each
slip system $s$ passes through this cell. In this case, positive and negative dislocations meeting
in this cell annihilate and the only relevant quantity is the excess density $n^s$. However, if the
mesoscopic cell comprises a number of unit cells, shown schematically in \reffig{fig_mesocell_size}b,
which is also the case in our model, several parallel slip planes pass through the cell. If we
consider that the dislocations move by pure glide, the mesoscopic cell can thus contain
simultaneously positive and negative dislocations in parallel slip planes and only a limited number
of dislocations can annihilate. Clearly, if we are to reproduce correctly the dislocation content of
the mesoscopic cell we have to consider the densities $n^{s+}$ and $n^{s-}$ separately.

Without the presence of dislocation sources, the densities $n^{s+}$ and $n^{s-}$ of crystal
dislocations in each slip system $s$ integrated through the simulated domain must be conserved
individually. This implies that the evolution equation for these dislocations densities must take
the form of a continuity equation. Since we consider here that crystal dislocations move only by glide
in their corresponding slip planes, the evolution equation for the dislocation densities reads
\begin{equation}
  \frac{\partial n^{s\pm}(\bsym{r},t)}{\partial t} = -D\ \nabla \cdot 
  \left[ \bsym{F}_{glide}^{s\pm}(\bsym{r},t) n^{s\pm}(\bsym{r},t) \right] \quad.
  \label{eq_FPeqn_sep}
\end{equation}
Here, $\bsym{F}_{glide}^{s\pm}(\bsym{r},t)$ represents the glide component of the Peach-Koehler
force on the crystal dislocations with the densities $n^{s\pm}$ in the mesoscopic cell at $\bsym{r}$
and time $t$, and can be calculated as follows. If the internal stress tensor is known at
$\bsym{r}$, the components of the total Peach-Koehler force on each crystal dislocation with the
Burgers vector $\pm{}\bsym{b}^s$ within the same mesoscopic cell can be calculated \cite{hirth:82,
  hull:01} as $F_k^{s\pm} = \mp\epsilon_{jk}\sigma_{jl}b_l^s$. The glide force is then determined by
projecting this force into the corresponding slip plane, i.e. $\bsym{F}_{glide}^{s\pm} =
(\bsym{F}^{s\pm} \cdot \bsym{e}^{s\pm})\bsym{e}^{s\pm}$, where $\bsym{e}^{s\pm} =
\bsym{b}^s/|\bsym{b}^s|$. The dislocation densities can then be updated using
\refeq{eq_FPeqn_sep}. For each slip system $s$ the fields $n^{s+}$ and $n^{s-}$ then enter
\refeq{eq_eta33_slipsys} to calculate the incompatibility field $\eta_{33}$. One can thus
recalculate the order parameter $e_2$ that minimizes the free energy \refeq{eq_freeE} subject to
this updated distribution of incompatibilities. The calculation is regarded as complete when the
free energy is minimized by the field $e_2$ and, simultaneously, the corresponding stress field does
not induce significant changes in the dislocation densities $n^{s+}$ and $n^{s-}$.  This process
thus represents a self-consistent procedure for the simultaneous evolution of the primary order
parameter field $e_2$ and of the densities of crystal dislocations in individual slip systems.


\section{Simulations}
\label{sec_simul}

The material considered in these simulations is the shape memory alloy single crystal of
Fe-30~at.\%Pd for which the temperature dependence of the elastic constants, measured by
\citet{muto:90}, is parametrized together with other coefficients entering the free energy
functional in Appendix~\ref{apx_FePdconst}. In our calculations the orientation of the crystal is
chosen such that the $x_1$ axis coincides with the $[100]$ direction, and the $x_2$ axis is parallel
to the $[010]$ direction.  Each mesoscopic cell is characterized by $1000\times 1000$
crystallographic unit cells with the lattice parameter $a_0 = 3.8$~\AA. Hence the width of each
mesoscopic cell is $a=0.38~\um$ and its area $S_{cell} \approx 0.14~\um^2$.  The initial values of
the order parameter $e_2$ in each mesoscopic cell are drawn randomly from a uniform distribution
with limits $\langle -0.04, 0.04 \rangle$ which include the values of $e_2$ that correspond to the
minima of $f_{loc}$ below $T_c$. The simulated two-dimensional domain contains $128 \times 128$
mesoscopic cells with periodic boundary conditions that are automatically imposed by the k-space
calculations. The width of the simulated domain is $48.64~\um$. For simplicity, no external stress
is applied in any simulation.  The minimization of the free energy is regarded as complete when
$\delta F/\delta e_2$ in \refeq{eq_e2evolve} becomes less than $0.001~\eV/$\AA$^3$.

The TEM observations of dislocations in polytwinned Fe-Pd thin foils due to \citet{xu:04} reveal
that the active mode accommodating plastic strain in this material is due to the glide of
$1/2\gdir{110}$ dislocations. Hence, in our two-dimensional simulations we will consider only two
slip systems $s$, populated by crystal dislocations with Bugers vectors
$\bsym{b}^{1\pm}=\pm{}1/2[110]$ and $\bsym{b}^{2\pm}=\pm{}1/2[\bar{1}10]$. Each mesoscopic cell is
initially assigned finite densities of these crystal dislocations, $n^{s+}$ and $n^{s-}$, each of
which is chosen at random from a uniform distribution between 0 and $n_{max}$, where $n_{max}$ is
varied to arrive at different dislocation densities. For each simulation the actual dislocation
density $\rho$ is given below.

\subsection{Martensitic texture as a function of the density of dislocations}

We first study the distribution of the order parameter $e_2$ that minimizes the free energy
\refeq{eq_freeE}: (i) in dislocation-free material, and (ii) subjected to fixed dislocation density.
Three finite densities of crystal dislocations are considered, given by
$\rho=\{6\times10^{14},10^{15},8\times10^{15}\}~\m^{-2}$. The free energy \refeq{eq_freeE} is then
minimized subject to the distribution of incompatibilities $\eta_{33}$ that are derived from the
given dislocation density using \refeq{eq_eta33_slipsys}. No evolution of the dislocation density is
allowed in this case which corresponds to a hypothetical situation where all dislocations are
immobile.  This minimization is performed for the temperatures both above $T_c$, where the austenite
is stable in the defect-free medium, and below $T_c$, where the martensite is the stable phase.

\begin{figure}[!htb]
  \begin{flushleft}
    $T>T_c$ \\
    \includegraphics[width=4cm]{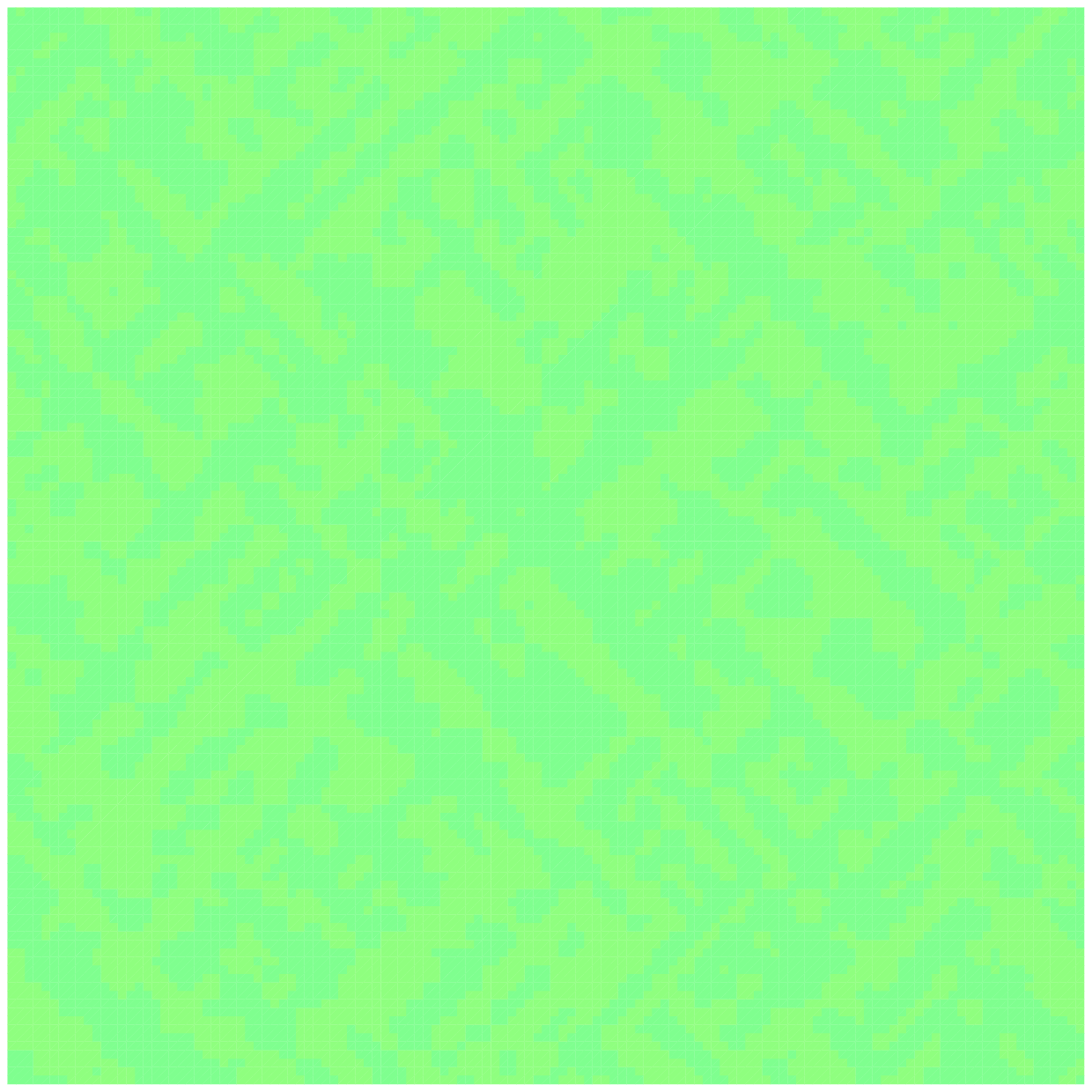}
    \includegraphics[width=4cm]{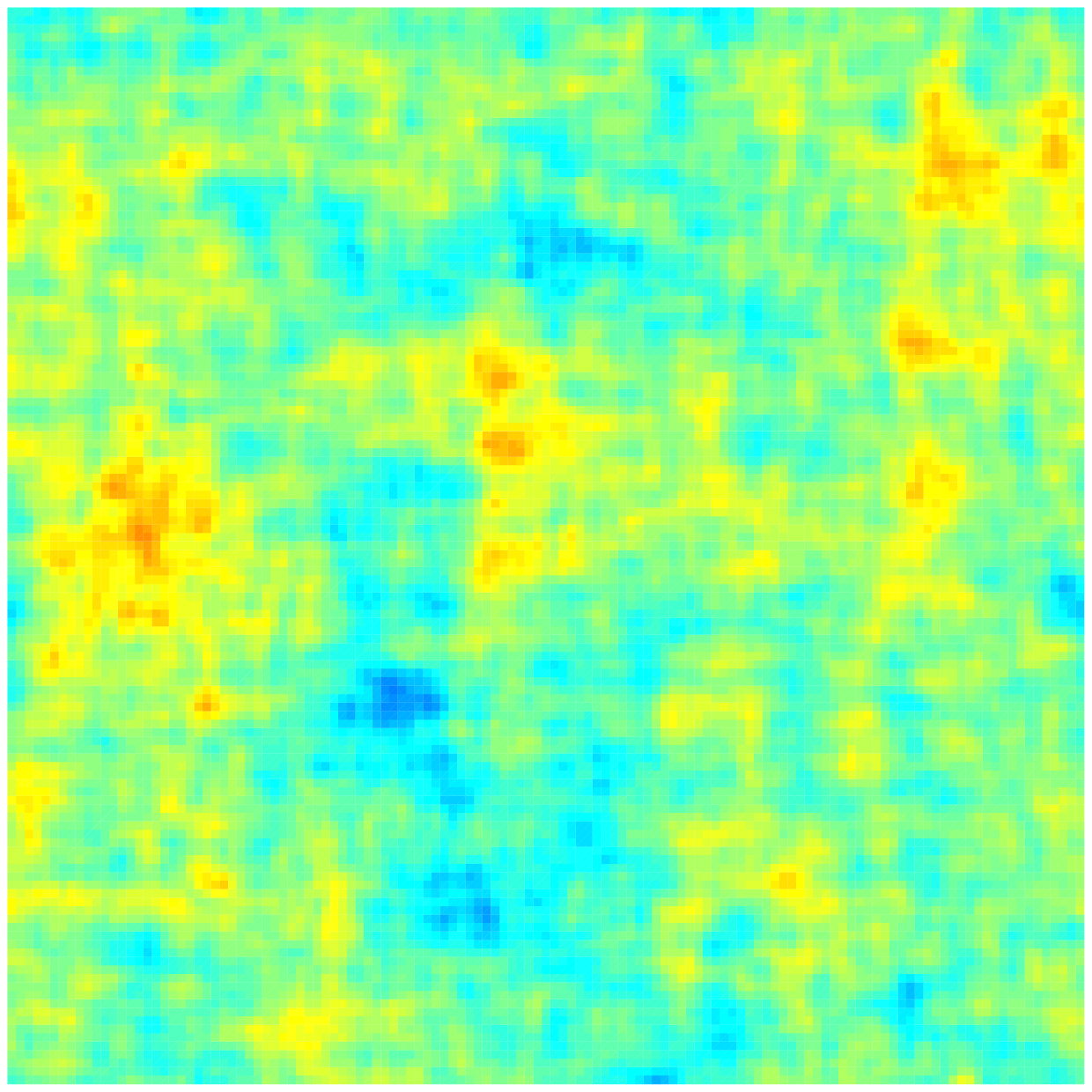}
    \includegraphics[width=4cm]{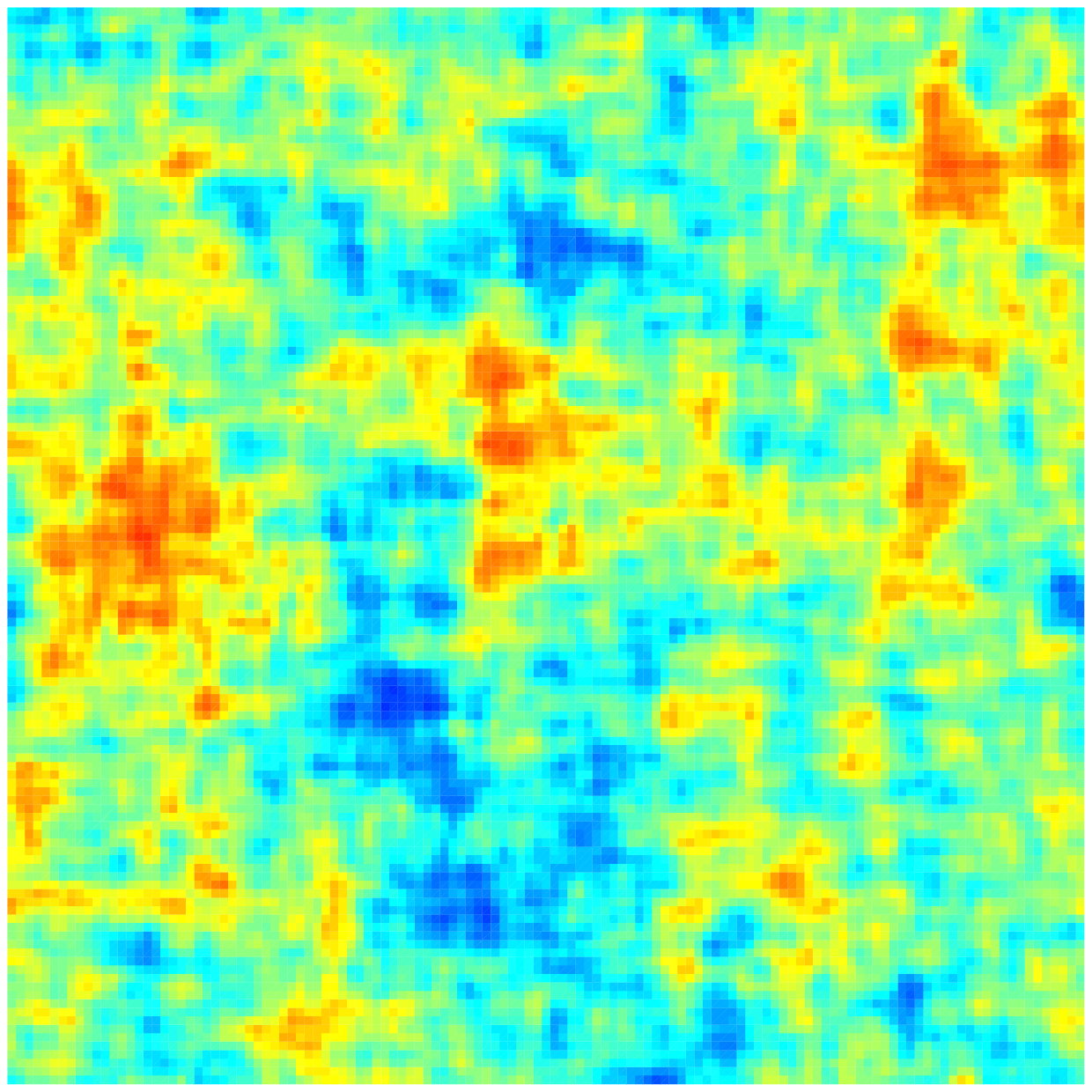}
    \includegraphics[width=4cm]{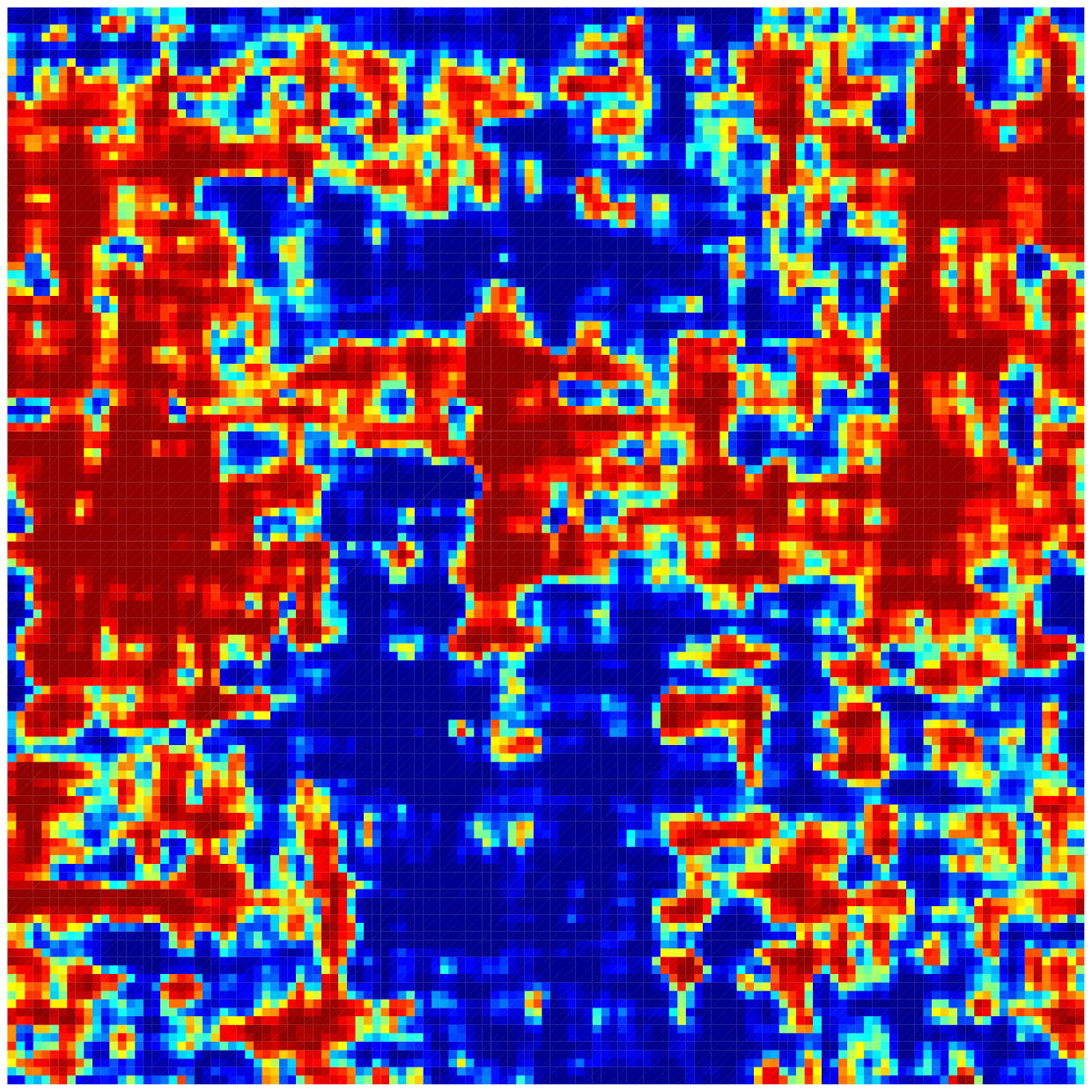}
  \end{flushleft} 
  \begin{flushleft}
    $T<T_c$ \\
    \includegraphics[width=4cm]{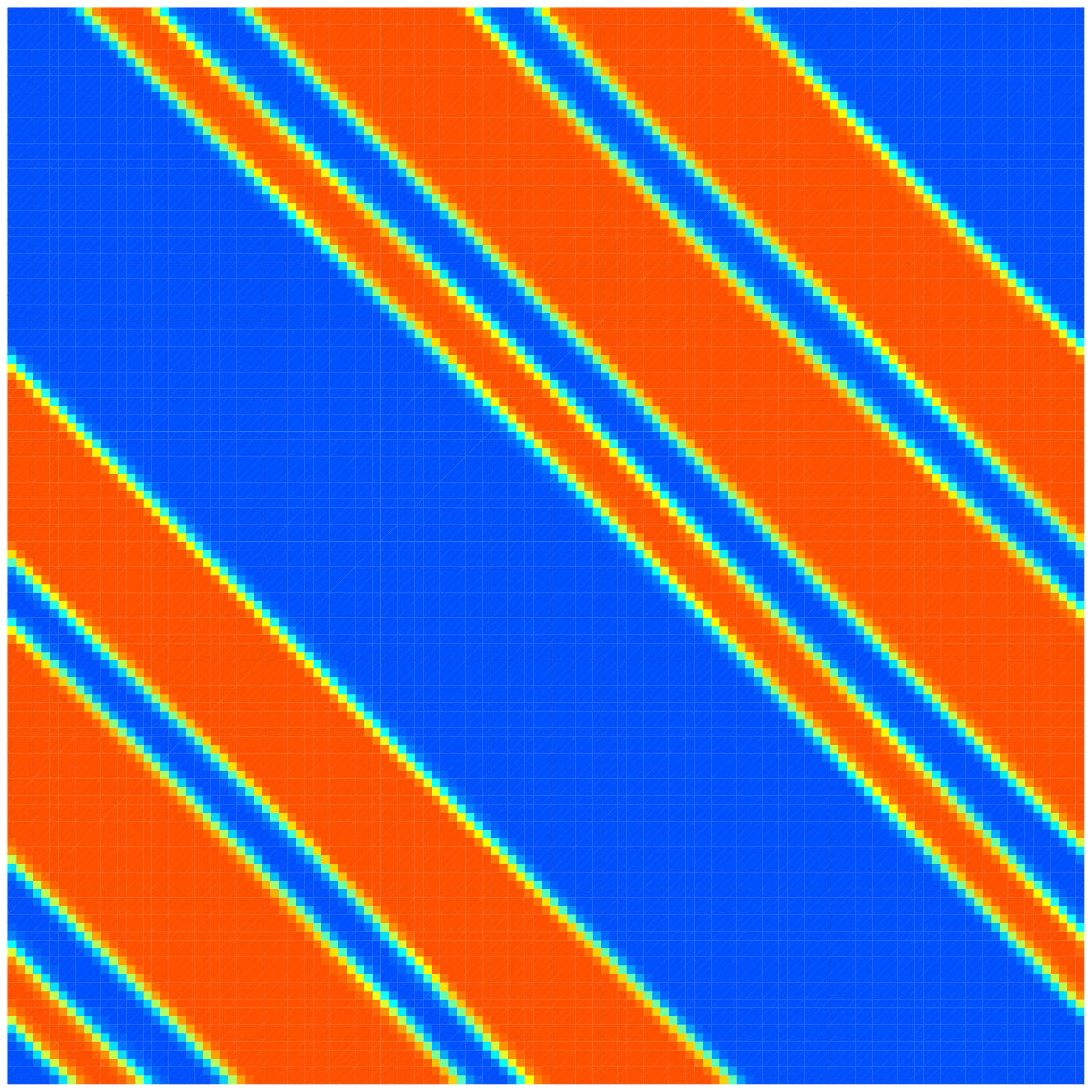}
    \includegraphics[width=4cm]{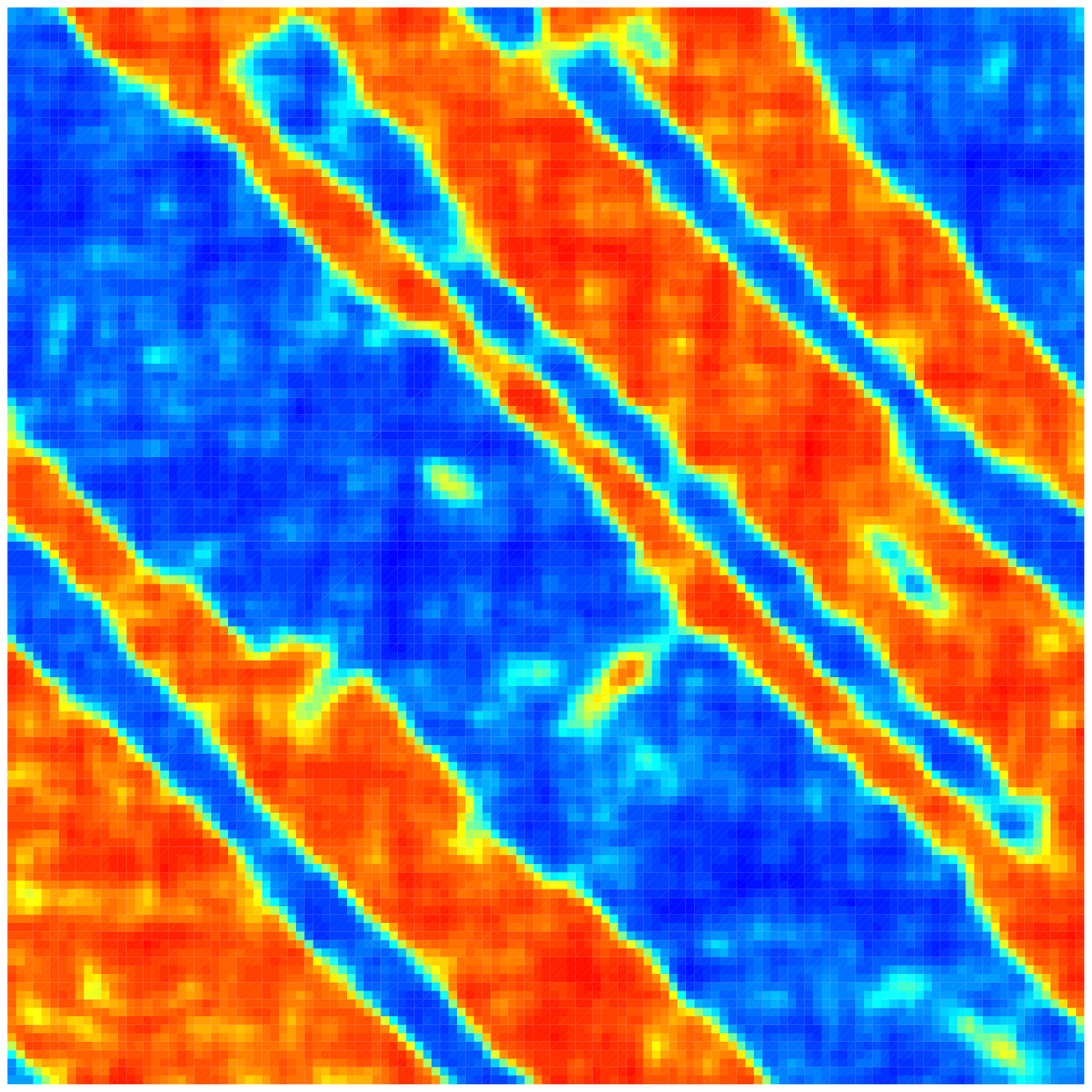}
    \includegraphics[width=4cm]{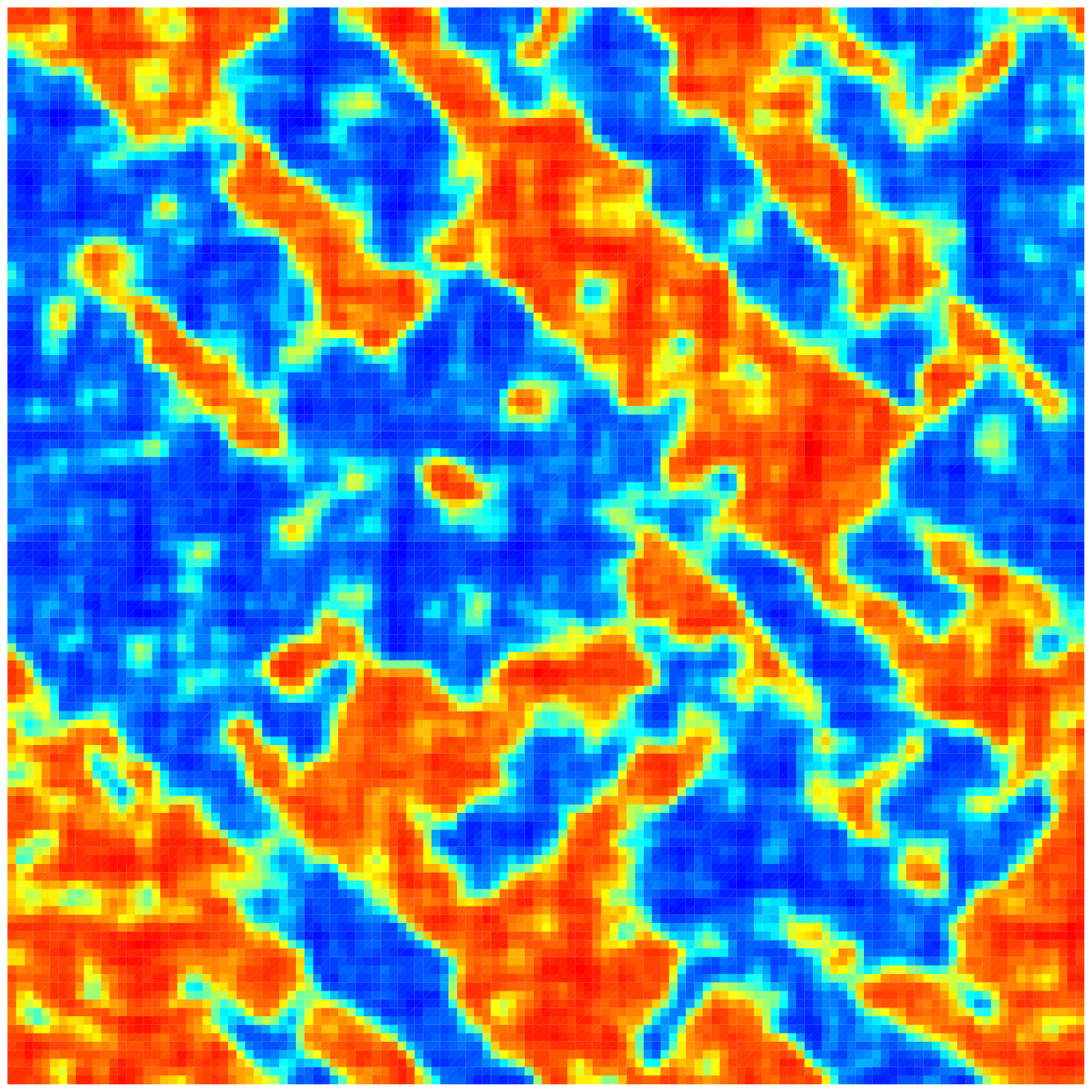}
    \includegraphics[width=4cm]{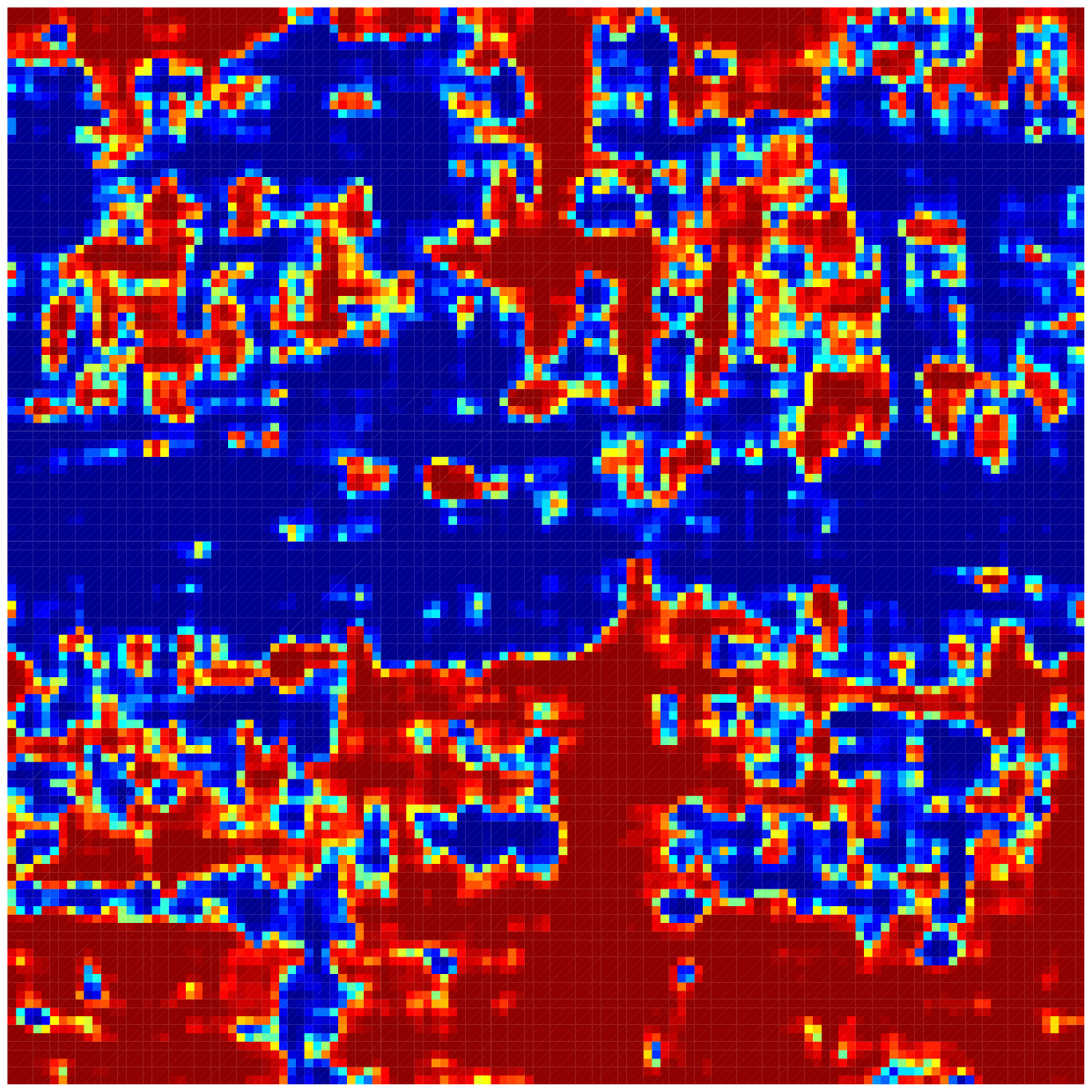} \\
    \centerline{(a) \hskip3.5cm (b) \hskip3.5cm (c) \hskip3.5cm (d)}
  \end{flushleft}
  \caption{Distribution of the order parameter field $e_2(\rr)$ above and below $T_c$ for
    dislocation-free medium as a function of the density of dislocations. (a) no dislocations, (b)
    $\rho = 6\times{}10^{14}\ \m^{-2}$, (c) $\rho = 10^{15}\ \m^{-2}$, (d) $\rho = 8\times{}10^{15}\
    \m^{-2}$.}
  \label{fig_e2_rho}
\end{figure}

In \reffig{fig_e2_rho} we show how the order parameter field $e_2$ that minimizes the free energy
\refeq{eq_freeE} changes as a function of the density of dislocations. The upper row corresponds to
the temperature above $T_c$ whereas the lower row to the temperature below $T_c$. Above $T_c$, the
system is progressively driven away from its free energy minimum ($e_2=0$) the higher the
dislocation density $\rho$. This results in stabilization of the martensitic phase above $T_c$ in
the order parameter field $e_2$ without a well-defined texture (\reffig{fig_e2_rho}b-d).  Below
$T_c$, the twinned microstructure that minimizes the free energy in dislocation-free materials
changes at finite densities of dislocations as a consequence of the competition of the first two
terms on the right-hand side of \refeq{eq_fnonloc_final}. For low densities $\rho$, the first term
dominates and the order parameter field $e_2$ is characterized by a twinned microstructure.  With
increasing dislocation density, the second term in \refeq{eq_fnonloc_final} becomes of the same
order as the first and this competition gradually causes elimination of the twins
(\reffig{fig_e2_rho}b-d). For large dislocation densities, i.e. \reffig{fig_e2_rho}d, the strain
incompatibility $\eta_{33}$ completely dominates the minimization of the free energy and the same
martensitic texture is obtained both above and below $T_c$ for the defect-free medium.  These
results are consistent with our previous observations that dislocations alter the character of the
martensitic texture (see \reffig{fig_frustr}).

Recall that the results shown in \reffig{fig_e2_rho} correspond to a highly idealized case where all
dislocations are immobile and the dislocation density thus cannot evolve in response to changes in
the order parameter field. In the following example, we remove this constraint to simultaneously
evolve the order parameter field and the dislocation density to provide a clearer picture of
dislocation pattern formation below $T_c$.

\subsection{Formation of dislocation walls at twin boundaries}

The initial distribution of the order parameter field above and below $T_c$ was obtained by
minimizing the free energy for a dislocation-free material. In each mesoscopic cell, the two
directions of the Burgers vectors in both slip systems $s$ were then assigned randomly a dislocation
density between 0 and $n_{max}=10^{14}~\m^{-2}$ which yields an actual dislocation density $\rho = 2
\times 10^{14}~\m^{-2}$. We then calculated the order parameter field that corresponds to the given
initial distribution of dislocations. This $e_2$ field is used to update the dislocation density
using \refeq{eq_FPeqn_sep} where all dislocations have the same mobility, i.e. $D$ is the same for
the evolution of all densities $n^{s\pm}$.  Since no external stress is applied the evolution of
the dislocation density is driven entirely by the evolving martensitic texture and the long-range
interactions between dislocations.

\begin{figure}[!htb]
  \centering
  \includegraphics[height=5cm]{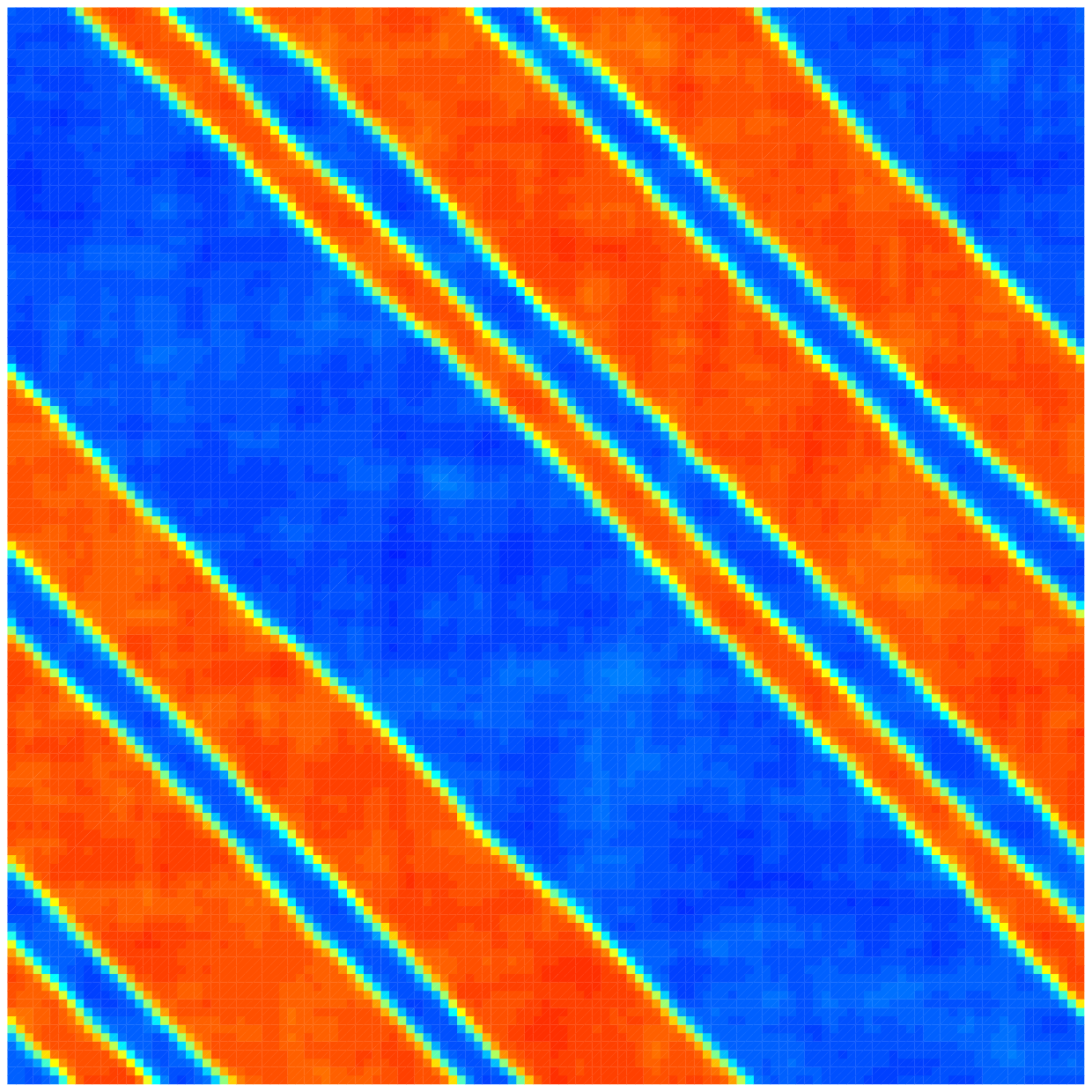} \qquad
  \includegraphics[height=5cm]{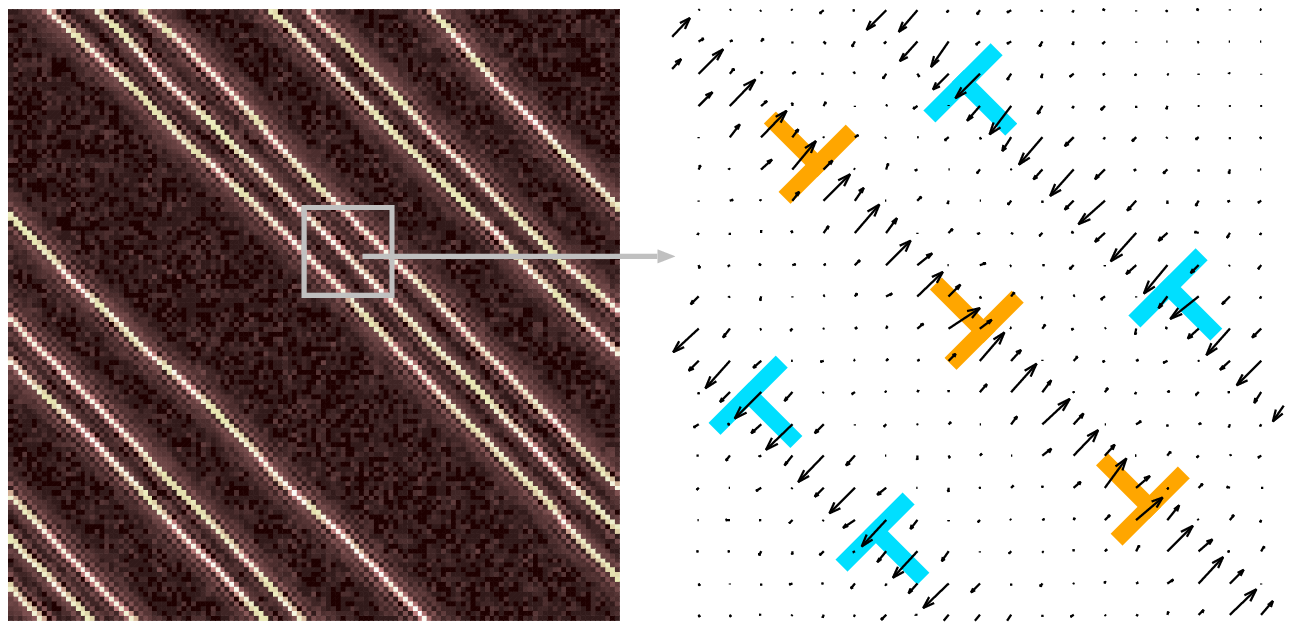} \\
  \hskip-2.5cm (a) \hskip8cm (b)
  \caption{The order parameter field $e_2$ (a) and the distribution of the magnitudes of net Burgers
    vectors $|\bsym{B}|$ (b) that minimize the free energy \refeq{eq_freeE}. The blue and
    red regions in (a) are the two variants of the martensite while the twin boundaries
    correspond to the austenite phase. In (b) the bright spots correspond to regions of high
    dislocation density and dark spots to low dislocation density. The inset of (b) shows the
    orientations of net Burgers vectors in the region marked by the gray square.}
  \label{fig_mart128}
\end{figure}

Above $T_c$ the order parameter field $e_2$ is close to zero everywhere and the corresponding
dislocation density is spatially uniform. Below $T_c$ the field $e_2$ is represented by a series of
twins corresponding to the two variants of martensite (see \reffig{fig_mart128}a). As can be seen
from \reffig{fig_mart128}b the scalar density of net Burgers vectors,
$|\bsym{B}(\bsym{r})|=\sqrt{B_1(\bsym{r})^2+B_2(\bsym{r})^2}$, corresponding to the minimum of the
free energy, is characterized by twin boundary dislocations with the Burgers vectors parallel to the
$[110]$ and $[\bar{1}\bar{1}0]$ directions. Hence, the twin boundaries are decorated by dislocation
walls with the net Burgers vectors parallel to the $\gdir{110}$ direction, as shown in the inset of
\reffig{fig_mart128}b. This results in the formation of alternating positive and negative
dislocation walls along the twin boundaries. Since all $\gdir{110}$ twins have the same energy, the
free energy can also be minimized by forming the $45^\circ$ (rather than $135^\circ$) texture in
$e_2$ (not shown here). In this case, the other two types of crystal dislocations, with Burgers
vectors $1/2[1\bar{1}0]$ and $1/2[\bar{1}10]$, would form the dislocation walls at the twin
boundaries.


\section{Conclusions}

The Landau-Ginzburg theory developed in this paper represents the first step in the formulation of a
mesoscopic model for martensitic phase transformations mediated by defects. In this framework the
presence of dislocations induces incompatibility between the components of the elastic strain tensor
and, therefore, the Saint-Venant condition no longer applies in media with defects.  Instead, the
elastic strains are related by an incompatibility constraint and the ``strength" of this
incompatibility is proportional to the gradients of the components of the Nye dislocation density
tensor\cite{nye:53}.  This incompatibility can be written in terms of densities of crystal
dislocations in individual discrete slip systems which provides a clear ``recipe" for coarse-graining
microscopic information to the mesoscopic description. The incompatibility field is then completely
determined and we have shown that its existence leads to such interesting phenomena as the
dependence of the order parameter texture on the density of dislocations or formation of correlated 
dislocations walls along the twin boundaries below $T_c$. 

The presence of dislocations, i.e. the existence of a finite incompatibility field, introduces new
contributions to the free energy functional, in particular coupling of the order parameter with the
incompatibility field and with the applied stress tensor. We have shown that the free energy minimum
cannot be reached by minimizing independently all terms that depend on the order parameter. During
this minimization the coupling between the order parameter and the incompatibility field introduces
competition between these terms and the minimization of the free energy is thus inherently
frustrated by the finite dislocation density. The order parameter field that minimizes the free
energy subject to a given distribution of dislocations (i.e. strain incompatibilities) can be
directly used to calculate internal strain and stress fields. For an arbitrarily anisotropic
material, one can thus obtain the Peach-Koehler forces on the dislocations just by minimizing the
free energy. These forces have been employed in the evolution equations for the density of
individual variants of crystal dislocations ($n^{s+}$ and $n^{s-}$) which takes the form of the
Fokker-Planck equation. Hence, the conservation of the total Burgers vector is automatically
satisfied.  This procedure represents a simple self-consistent scheme in which the order parameter
field is first calculated by minimizing the free energy subject to a given distribution of strain
incompatibilities, i.e. densities of crystal dislocations. The corresponding internal stress field
and the Peach-Koehler forces are then used to update the dislocation density using the discretized
version of the Fokker-Planck equation.  The new distribution of strain incompatibilities that
correspond to the updated dislocation density is used again to minimize the free energy and this in
turn provides the new order parameter field.

To demonstrate the main features of the model developed in this paper, we considered two case
studies that represent the limits of the theory with regard to the mobility of the dislocations. In
the first case the four crystal dislocations with the $1/2\gdir{110}$ Burgers vectors were
considered as immobile and, therefore, the free energy was subject to a fixed distribution of strain
incompatibilities and minimized purely by the order parameter field $e_2$. We have shown that the
nucleation of martensite and its spatial distribution depends on the density of dislocations. At low
dislocation densities the minimization of the free energy is only weakly affected and the
corresponding microstructure below $T_c$ is represented by martensite twins along the $\gdir{110}$
directions. In contrast, at high dislocation densities the distribution of strain incompatibilities
strongly affects the minimization of the free energy and eliminates the twinned microstructure. In
the second case study we considered that all dislocations are completely mobile. We have
demonstrated that below $T_c$ the free energy is minimized by forming alternating net dislocation
walls at the mesoscale that decorate the twin boundaries between different variants of the
martensite. The results are qualitatively consistent with the observations of correlated dislocation
domains along twin boundaries in Ag\citep{worzala:67}, Ni-Ti\cite{sehitoglu:03} and
Fe-Pd\cite{halley:02}.

Our objective here was mainly to demonstrate how continuum theory of dislocations\cite{kroner:58}
can be incorporated into the mesoscopic free energy functional for displacive phase
transformations\cite{kartha:95} which inevitably led to a number of simplifications. Many
crystallographic details are beyond the resolution of the model and thus are included only in a
coarse-grained manner. In particular, short-range interactions between crystal dislocations are
replaced by continuous dislocation density fields $n^{s+}$ and $n^{s-}$ that correspond to the two
directions of the Burgers vector in the slip system $s$. The individual slip systems are assumed to
be independent of each other and no reactions between dislocations moving in different slip planes
are incorporated. No upper limit on the density of dislocations due to a finite minimum distance
between dislocations of the same type is imposed at this point. Since no external loading is applied
here, we do not consider the existence and operation of dislocation sources.

In the future, the model developed in this paper will be advanced to study the effect of
dislocations on strain hardening and hysteresis, both of which play crucial roles in shape memory
alloys. It serves as the basis for analytical solutions of elastic fringing fields at habit planes
in the presence of dislocations and thus allows for a study of the role of defects on the size
dependence of the twinning width in martensites. The concept of the defect-induced incompatibility
of elastic strains is general and the approach pioneered in this paper may be applied to other phase
transitions that are mediated by defects. Examples include the study of strain-induced polarization
in ferroelectrics, strain-induced magnetization in ferromagnets or even the effect of long-range
strain fields of dislocations on the recently discovered ferrotoroidal ordering\cite{vanaken:07} in
ferrotoroidic materials.


\begin{acknowledgments}
  The authors thank F.-J.~P\'erez-Reche, R.~Ahluwalia, K.~Dayal, S.~Sengupta, J.~San Juan and
  A.~Roytburd for their comments and stimulating discussion of this work during the International
  Conference on Martensitic Transformations (ICOMAT-08) in Santa Fe, New Mexico. In addition, we
  thank A.~Acharya for bringing to our attention his work on the subject.
\end{acknowledgments}


\appendix
\section{Tensorial representations}
\label{apx_tensors}

For convenience, following are tensorial representations of the operations that are used throughout
this paper\cite{chandrasekharaiah:94}. Here, $\epsilon_{ijk}$ is the antisymmetric Levi-Civita
tensor and $\bsym{A}$ a tensor of rank two.\\

\noindent
\begin{tabular}{r@{\hskip1em}l}
  divergence: & $\nabla \cdot \bsym{A} = A_{ij,i}$\\
  curl: & $\nabla \times \bsym{A} = \epsilon_{imn} A_{jn,m}$ \\
  symmetric curl: & $\sym (\nabla \times \bsym{A}) = \frac{1}{2}(\epsilon_{imn}A_{jn,m}+\epsilon_{jmn}A_{in,m})$ \\
  incompatibility:  & $\nabla \times \nabla \times \bsym{A} = \epsilon_{irs}\epsilon_{jmn}A_{sn,mr}$
\end{tabular}

\section{k-space kernels}
\label{apx_kernels}

In the following we write explicitly all the k-space kernels that are used to calculate the free
energy and to perform the relaxational dynamics for the order parameter $e_2$. The denominators in
these kernels are identical and we evaluate them separately as
\begin{equation}
  d(\k) = \frac{1}{A_1} k^4 + \frac{8}{A_3} k_x^2 k_y^2 \quad.
\end{equation}
The following kernels are used to calculate the secondary order parameter fields $e_1$ and $e_3$
from the known primary order parameter field $e_2$, the incompatibility field $\eta_{33}$, and the
external stress field $\sigma_{ij}$:
\begin{eqnarray}
  \nonumber
  Q_1(\k) = \frac{1}{A_1} \frac{k_x^4-k_y^4}{d(\k)} \quad,\quad
    Q_3(\k) = -\frac{\sqrt{8}}{A_3} \frac{k_xk_y(k_x^2-k_y^2)}{d(\k)} \quad,\\
  \nonumber
  R_1(\k) = \frac{1}{A_1} \frac{k^2}{d(\k)} \quad,\quad
    R_3(\k) = -\frac{\sqrt{8}}{A_3} \frac{k_x k_y}{d(\k)} \quad,\\
  S_1(\k) = \frac{1}{A_1\sqrt{8}} \left[ \frac{1}{A_1}\frac{k^4}{d(\k)} + 1 \right] \quad,\quad
    S_3(\k) = -\frac{1}{A_1A_3} \frac{k_x k_y k^2}{d(\k)} \quad,\\
  \nonumber
  T_1(\k) = -\frac{\sqrt{8}}{A_1A_3} \frac{k_x k_y k^2}{d(\k)} \quad,\quad
    T_3(\k) = \frac{1}{A_3} \left[ \frac{8}{A_3}\frac{k_x^2 k_y^2}{d(\k)} - 1 \right] \quad.
\end{eqnarray}

The nonlocal part of the free energy density, $f_{nonloc}$, and its partial derivative, $\partial
f_{nonloc}/\partial e_2$, are expressed using
\begin{equation}
  A_{13}(\k) = \frac{(k_x^2-k_y^2)^2}{d(\k)} \quad,\quad
  B_{13}(\k) = \frac{k_x^2-k_y^2}{d(\k)} \quad,\quad
  C_{13}(\k) = \frac{1}{d(\k)} \quad,
\end{equation}
and the kernels that depend on the above and the applied stress tensor are
\begin{eqnarray}
  \nonumber
  \Sigma_{AQ}(\k) &=& A_1 Q_1(\k) m_1(\k) - 
    A_3 Q_3(\k) m_3(\k)  \quad,\\
  \Sigma_{AR}(\k) &=& A_1 R_1(\k) m_1(\k) - 
    A_3 R_3(\k) m_3(\k) \quad,\\
  \nonumber
  \Sigma_A(\k) &=& A_1 [m_1(\k)]^2 + A_3 [m_3(\k)]^2 \quad,
\end{eqnarray}
where:
\begin{eqnarray}
  \nonumber
  m_1(\k) &=& S_1(\k)\sigma_{jj}(\k) - T_1(\k)\sigma_{12}(\k) \quad,\\
  m_3(\k) &=& S_3(\k)\sigma_{jj}(\k) - T_3(\k)\sigma_{12}(\k) \quad,
\end{eqnarray}
Here, the subscripts of each $\Sigma$ imply which combination of the coefficients ${A_i, Q_i, R_i}$
is used, and $\sigma_{jj} = \sigma_{11} + \sigma_{22}$.

Finally, the strain energy density, $f_{load}$, is expressed using the kernels that depend only on
the applied stress tensor, $\sigma_{ij}$, as:
\begin{eqnarray}
  \nonumber
  W_Q(\k) &=& \frac{1}{\sqrt{8}} \left\{ \left[ Q_1(\k)+1 \right] \sigma_{11}(\k) + 
    \left[ Q_1(\k)-1 \right] \sigma_{22}(\k) \right\} + 
    Q_3(\k) \sigma_{12}(\k) \quad,\\
  W_R(\k) &=& \frac{1}{\sqrt{8}} R_1(\k) \sigma_{jj}(\k) + R_3(\k)\sigma_{12}(\k) \quad,\\
  \nonumber
  W_\Sigma(\k) &=& \frac{1}{\sqrt{8}} \sigma_{jj}(\k) m_1(\k) - \sigma_{12}(\k) m_3(\k) \quad.
\end{eqnarray}

\section{Temperature dependence of the elastic constants for Fe-30at.\%Pd}
\label{apx_FePdconst}

\begin{figure}[htb]
  \centering
  \includegraphics[width=11cm]{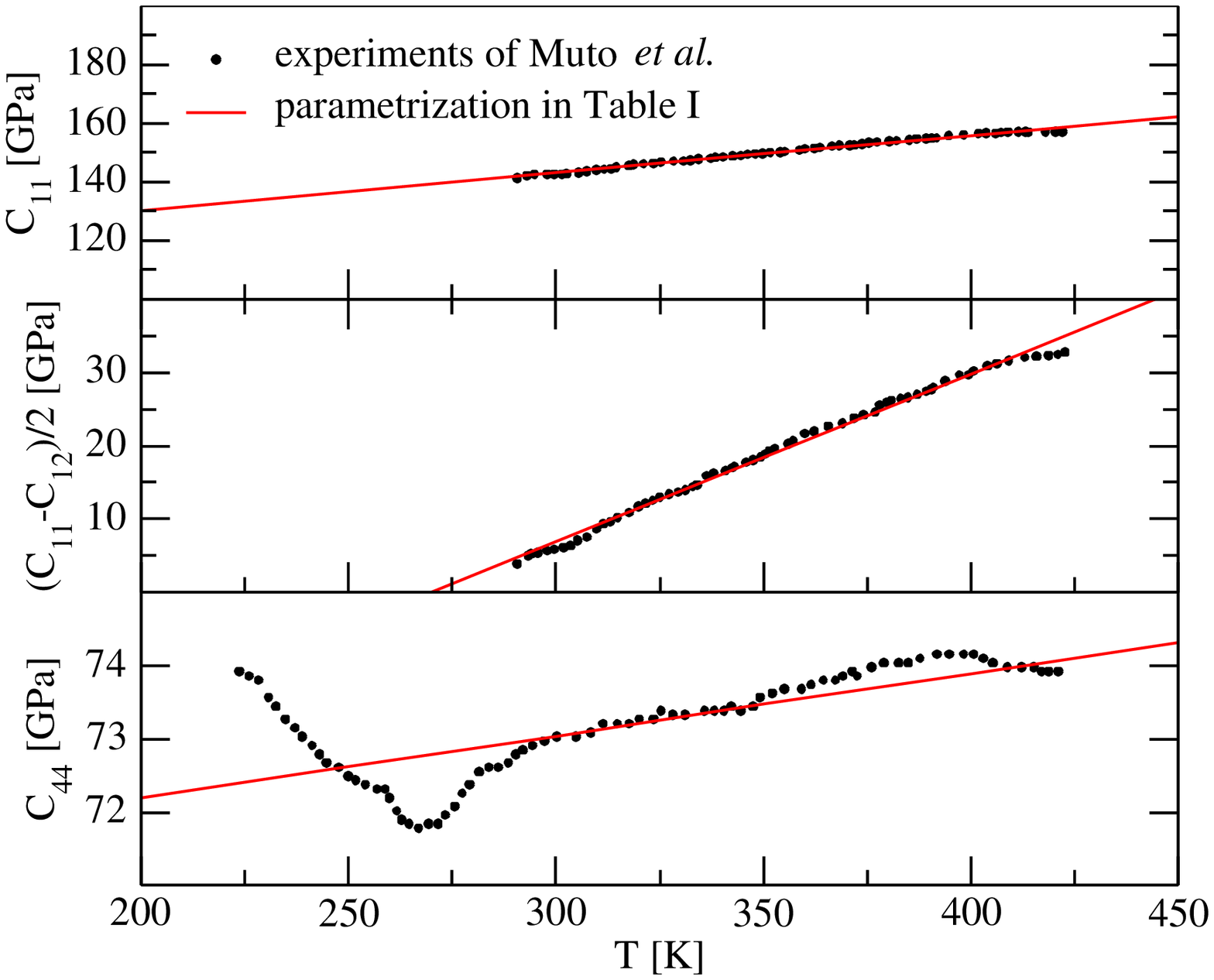}
  \caption{Temperature scaling of elastic constants for Fe-30at.\%Pd from \reftab{tab_FePdconst}
    compared with the experimental data of \citet{muto:90}.}
  \label{fig_FePdcomp}
\end{figure}

The temperature dependence of the elastic constants of Fe-30at.\%Pd alloy, measured by
\citet{muto:90} and shown in \reffig{fig_FePdcomp}, is approximated by linear relations given in
\reftab{tab_FePdconst}. The scaling laws are obtained by fitting the experimental data for $C_{11}$,
$(C_{11}+C_{12}+2C_{44})/2$ and $(C_{11}-C_{12})/2$ that are linear within the range of temperatures
considered in these measurements. The remaining parameters $B$, $C$, and $K_2$ are taken from
Ref.~\onlinecite{kartha:95} and are assumed to be independent of temperature: $B = -1.7\times{}10^4\ \GPa$, $C
= 3\times{}10^7\ \GPa$, $K_2/a^2 = 25\ \GPa$.

\begin{table}[!htb]
  \caption{Temperature scaling of the elastic constants and the coefficients used in the
    free energy \refeq{eq_freeE}. Units: $T$ [K], $C_{ij}$ [GPa], $A_i$ [GPa].}
  \label{tab_FePdconst}
  \begin{tabular}{ccccc}
    \hline
    $C_{11}(T)$ && $C_{12}(T)$ && $C_{44}(T)$ \\
    \hline
    $104.82 + 0.1279T$ && $229.09 - 0.3321T$ && $71.73 + 0.0010T$ \\
    \hline
  \end{tabular}
  \vskip1em

  \begin{tabular}{ccccc}
    \hline
    $A_1(T)$ && $A_2(T)$ && $A_3(T)$ \\
    \hline
    $333.91 - 0.2042T$ && $-124.27 + 0.4600T$ && $282.00 + 0.0344T$ \\
    \hline
  \end{tabular}
\end{table}

\newpage
\bibliography{bibliography}

\end{document}